\documentclass[usenatbib]{mn2e}
\usepackage{epsfig}
\usepackage{amsmath}
\usepackage{amssymb}
\usepackage{journal_names}
\usepackage{color}
\usepackage{multirow}

\begin{document}

\title{Wide-field imaging of NGC 4365's globular cluster system: The third subpopulation revisited}
\author{Christina Blom, Lee R.\ Spitler and Duncan A.\ Forbes}
%\footnotetext{Centre for Astrophysics \& Supercomputing, Swinburne University, Hawthorn VIC 3122, Australia}
\date{\today}
\maketitle

\begin{abstract}
Analysis of the globular cluster (GC) system of the giant elliptical (E3) galaxy NGC 4365, from eight Hubble Space Telescope/Advanced Camera for Surveys (ACS) pointings and a wide-field Subaru/SuprimeCam (S-Cam) image, is presented. Using magnitude, colour and size criteria we obtain a catalogue of GC candidates. We also measure the photometric properties of the galaxy starlight, including a new measure of the effective radius.%Fitting a S\'{e}rsic profile to the galaxy starlight we find the effective radius of the galaxy to be $2.1$ arcmin (14 kpc). Candidate GCs from S-Cam imaging are combined with those from eight Hubble Space Telescope/Advanced Camera for Surveys (HST/ACS) pointings around NGC 4365. 
We determine the lower limit on the number of GCs to be $6450\pm110$ and show that the GC system extends beyond 134 kpc (9.5 galaxy effective radii). We revisit the question of whether NGC 4365 has a bimodal or trimodal GC colour distribution and find support for three distinct GC colour subpopulations (i.e.\ blue, green and red). % when the GC system is restricted to the central 2 arcmin (13 kpc). %and hence, whether there are two or three distinct GC subpopulations. We do not find conclusive evidence for a trimodal colour distribution when the entire radial range of our imaging is used but we do find support for three distinct GC subpopulations when the properties of the blue, red and intermediate colour (green) GCs are compared. 
S\'{e}rsic profile fits to the radial surface density of each subpopulation reveal that the blue GCs are more extended than either red or green GCs. %and that the green subpopulation is only significant in the inner 2 arcmin (13 kpc). 
The median half light radii for GCs in the blue, green and red subpopulations are $4.1\pm^{0.3}_{0.2}$, $3.0\pm^{0.2}_{0.1}$ and $2.8\pm^{0.1}_{0.1}$ parsec respectively. The estimated subpopulation ellipticities are $0.66\pm0.06$, $0.55\pm0.07$ and $0.16\pm0.25$ for the blue, green and red GCs, where alignment with the photometric position angle of the galaxy ($\sim42^{\circ}$) is assumed. A KS test on the mass functions show a $>98$ per cent probability that all three subpopulations are distinct from one another. We also find radial gradients of GC size and colour (metallicity) and a blue tilt. %In the trimodal case the metallicity slopes with galactocentric radius are $-0.13\pm0.03$ and $-0.26\pm0.06$ dex per dex for the blue and red subpopulations respectively. %We conclude that it is likely that NGC 4365's GC system contains three distinct subpopulations and that the green subpopulation is contained within the central 13 kpc. 
The properties, including surface density profile, position angle, ellipticity and radial colour gradient, of the red GC subpopulation are very similar to the properties of NGC 4365's starlight. This result supports the hypothesis that red GCs are formed along with the bulk of the diffuse starlight in the galaxy. NGC 4365 has a kinematically distinct core and a significant misalignment between the photometric and kinematic major axes. %There are several explanations for the cause of these kinematic features and w
We discuss the possibility that these kinematic features are related to the presence of the distinct third GC subpopulation. %The most likely cause of the kinematic features, a combination of axisymmetric stellar orbits in the inner parts and triaxial orbits in the outer parts, is not obviously connected to the formation of a third GC subpopulation. 
We briefly discuss implications for the formation of NGC 4365, finding that major merger, multi-phase collapse and accretion formation scenarios could all account for the existence of a third GC subpopulation.
\end{abstract}
%\onecolumn
\section{Introduction}
%Globular clusters are among the oldest objects in the Universe and, as they have remained virtually unchanged since their formation, they are an important tool with which to study the evolutionary history of their host galaxies. By studying the colour multimodality, spatial distribution and size properties of GC systems, it is possible to investigate the evolutionary history of their host galaxies. 
%Observations of extragalactic GC systems allow us to do galactic archaeology for external galaxies. GCs have sizes ranging between 2 and 20 pc in diameter \citep*{vdb91} and colours ranging from 0.5 to 1.9 in g-z \citep{VCS16}. 
%Because GC systems extend much further than the diffuse stellar light, CCD imaging with a field-of-view wide enough to analyse the two dimensional structure of GC systems are rare. Most GC systems are bimodal in colour \citep{Br06} and there is evidence that GC properties such as size, radial and azimuthal distributions are linked to this bimodality.
\subsection{Globular cluster colour bimodality}
Bimodality in the colour distribution of GC systems in elliptical galaxies was first clearly demonstrated by \citet{Ze93}. Since then, it has been shown to be a common feature in large galaxies of all Hubble types (e.g. \citealt*{F97}; \citealt{La01}; \citealt{VCS9}). Spectroscopic observations of small samples of extragalactic GCs have generally found them to be very old ($\sim$ 12 Gyrs), similar to their Milky Way counterparts. Such old ages support the interpretation of colours as a good proxy for metallicity. Thus the photometric observations of large samples of GCs indicate two metallicity subpopulations: a blue, or metal poor, subpopulation and a red, or metal rich, subpopulation. These metallicity subpopulations often reveal distinct kinematic distributions \citep{Le10a,Ar11,Fos11} similar to those observed in the Milky Way's GC system \citep{Zi85}. The mean colour/metallicity of the two subpopulations correlate with the host galaxy luminosity \citep{VCS9}. The presence of distinct subpopulations indicates that GC system formation was not a simple one-step process with (at least) two phases of GC formation suggested. Various models have been proposed to explain the two formation modes \citep{As92,F97,C98,Be02,Be08}. %Recently, \citet{Mu10} have modelled the formation of GCs without two distinct formation modes however their simulated Milky Way GC system has an age-metallicity relation that does not match that observed for our Galaxy.

The interpretation of colour bimodality corresponding to metallicity bimodality has been challenged recently by \citet{Y06}, \citet{Ca07} and \citet{Bl10} who suggested that the transformation from colour into metallicity is strongly non-linear. Beyond the Milky Way, there are very few GC systems observed with reasonable numbers of GCs in which to examine the spectroscopic metallicity distribution directly. However for those that do exist, i.e. NGC 4472 \citep{St07}, NGC 5128 \citep{Wo10} and NGC 4594 \citep{AB11} spectroscopic metallicity bimodality is confirmed. 

With large photometric samples one can also examine other properties of extragalactic GC subpopulations. These include their radial and azimuthal distributions, luminosity or mass functions and physical sizes (if imaging with suitable resolution is available). Such properties also often reveal differences between the blue and red subpopulations \citep[see review by][]{Br06}. For example, the red GCs are usually more centrally concentrated than the blue GCs, and have a density profile, mean colour and spatial distribution that matches the bulge/spheroid component of the galaxy \citep{Fo01}. The blue GCs have been measured to have larger mean sizes on average than the red GCs \citep{Jo04}. 

Although most large galaxies with well-studied GC systems reveal clear colour bimodality, there are some exceptions \citep[see selected galaxies in][]{VCS9}. One notable exception is the giant elliptical galaxy NGC 4365 which reveals a broad colour distribution at all GC magnitudes. Here we revisit the issue of a possible third GC subpopulation at intermediate colours in NGC 4365.

\subsection{GC subpopulations of NGC 4365}
\begin{figure*}\centering
 \includegraphics[width=1.0\textwidth]{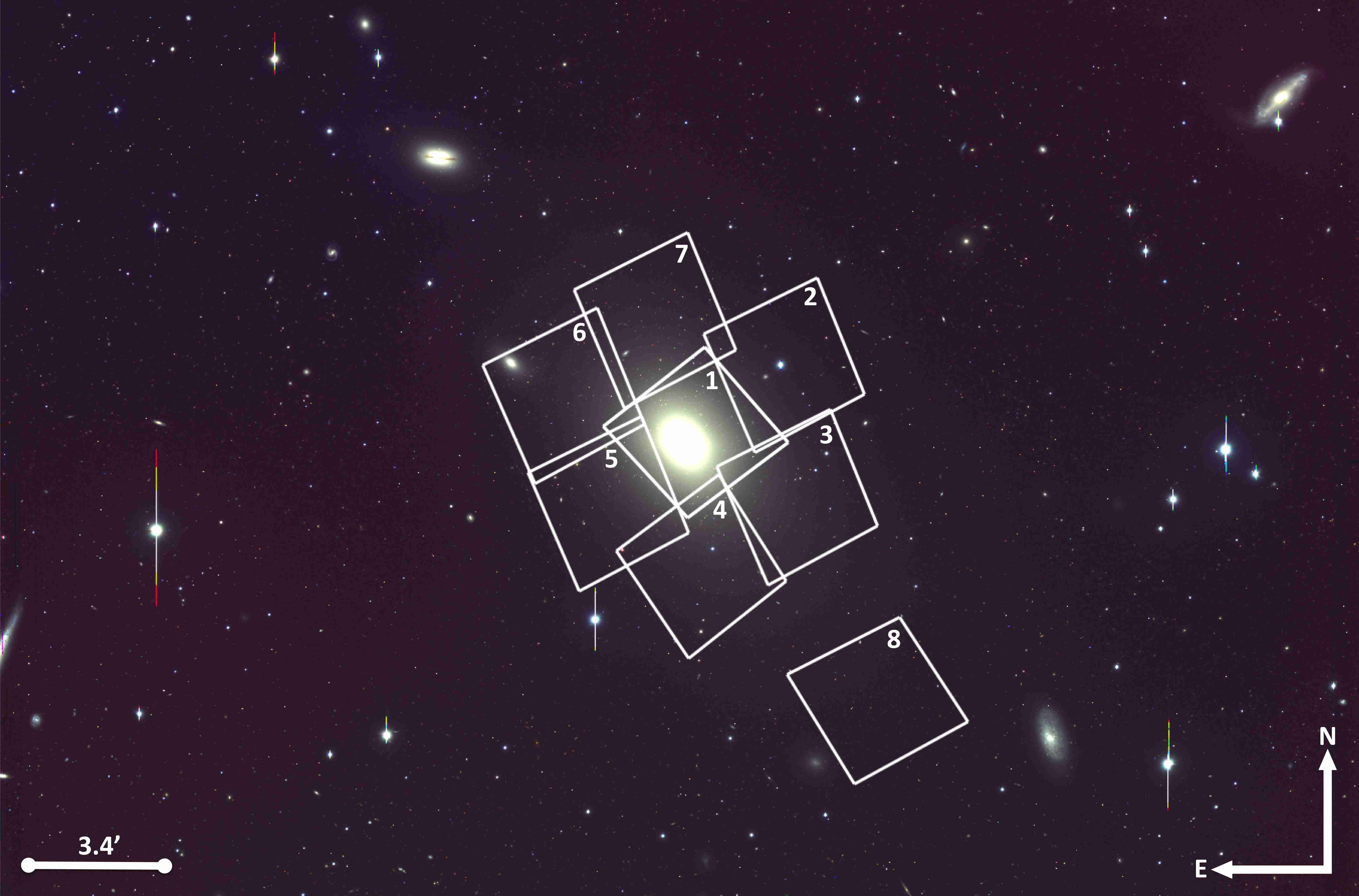}								%use footprintsplus.png for smaller pdf
 \caption{The central $34 \times 25$ arcmin section of the Subaru/S-Cam $g'$, $r'$ and $i'$ filter combined image (trimmed from $35 \times 27$ arcmin for cosmetic purposes) and the footprints of the 8 HST/ACS pointings are shown. The scale of the ACS pointings is shown in the bottom left corner. Image is centered on $\alpha$=12:24:26.824; $\delta$=+07:19:03.52 (J2000.0). At a distance of $23.1\pm0.8$ Mpc \citep{FCS5} $1 \mathrm{  arcsec} =0.112$ kpc.}
 \label{fig:footp}
\end{figure*}
Here we study NGC 4365, a giant elliptical (E3) galaxy \citep{VCS6} on the far edge of the Virgo Cluster, 23.1 Mpc away \citep{FCS5} with $M_B=-21.3$ mag \citep{VCS6}. There is debate as to whether the apparent Kinematically Distinct Core (KDC) of the galaxy is an actual kinematic property or the observed effect of triaxial orbits \citep[e.g.][]{Da01,vdB08}. NGC 4365 is also one of few galaxies that rotate about the photometric minor rather than major axis \citep{Da01}. %but \citet{vdB08} provide fairly convincing evidence that the latter is in fact true. %NGC 4365 has been studied extensively and it forms part of many surveys such as the SAURON survey and the ACS Virgo Cluster Survey. This galaxy shows no clear morphological evidence of interaction or recent merger, although 
\citet{Tal09} claim evidence of a faint fan feature in the SW of the galaxy. 

\citet{F96} and \citet{F96b} studied the GC system properties of NGC 4365 and other giant elliptical galaxies with KDCs. Bimodality was not suspected at the time in most GC systems and they found no significant differences between NGC 4365 and other giant elliptical galaxies with KDCs. Two independent groups working with HST/WFPC2 imaging, \citet{La01} and \citet{Ku01a}, found that the GC system of NGC 4365 was best fit by a unimodal distribution, whereas most of the elliptical galaxies in their samples %(including many of the \citet{F96b} galaxies) 
were best fit by bimodal distributions. Soon afterwards, \citet{Pu02} analysed the GC system of NGC 4365, using the combination of  HST/WFPC2 optical and VLT/ISAAC K-band photometry to break the age-metallicity degeneracy, which is a problem inherent to photometric analysis of GCs. In their analysis they found what appeared to be a slightly younger and very metal rich subpopulation in addition to two old subpopulations.  When \citet{La03} published Lick indices (obtained from Keck/LRIS) of 14 GCs in NGC 4365 it still looked likely that the galaxy hosted some old GCs and some very metal rich, young GCs that ``conspire to produce the single broad distribution observed in optical colors" \citep{La03}. 

However, when \citet{Br05} extended the sample of NGC 4365 GCs with Lick indices by adding 19 new objects \citep[also re-observing 3 objects from][]{La03} they found a uniform old age for all the GCs in the sample and suggested that the third GC subpopulation in the system, also found at intermediate optical colours, was due to a population with intermediate metallicity not young age. \citet{He04} published ``better age determinations" for the intermediate GC subpopulation by including observations in another infrared filter and \citet{Ku05} approached the problem with deeper HST observations in the H-band. \citet{La05} confirmed the presence of intermediate optical colour GCs at small galactocentric radii (from an analysis of optical HST/ACS photometry) and compared photometric age determinations with spectroscopic ones to assess the accuracy of photometrically determined ages. This work cast some doubt on the accuracy with which ages and metallicities can be measured using a combination of infrared and optical photometry. Later, \citet{He07} published a photometric comparison of elliptical galaxies in the centres of clusters and those in smaller groups, claiming (from optical and near-infrared photometry) that many group elliptical galaxies (including NGC 4365) host intermediate age GCs. Using g, z and K filter photometry, \citet{Ch11} showed that NGC 4365's GCs all have similar ages (much like the other large ellipticals in their sample) but that the distribution in the g-z direction was significantly different to other large ellipticals. %The debate has not been settled. While age resolution of photometric data is not as good as can be obtained in spectroscopic data it might still be able to resolve age-subpopulations.

While there is still debate on the nature (in age, colour and metallicity) of a third GC subpopulation in NGC 4365, there is consensus that the GC system of NGC 4365 is different to other galaxies of similar luminosity. %and requires an explanation. 
Most analyses of NGC 4365's GC system subpopulations to date have been done with tens or hundreds of GCs. We revisit this issue in colour using thousands of GC candidates from optical photometry over most of the spatial extent of the galaxy, to analyse whether there is statistically significant indication of three colour subpopulations in its GC system. %Once the significance and parameters of any GC subpopulations are determined, our data set makes it possible to characterise those subpopulations in terms of their azimuthal properties, size, radial extent, colour and metallicity (if old ages are assumed).

\subsection{Paper Structure}
In Section 2 and 3 we overview the data acquisition and reduction for the Subaru/S-Cam and HST/ACS photometry respectively. We describe the GC candidate selection criteria and detail the photometric properties of the GC candidate samples in Section 4. Section 5 describes the properties of the galaxy light and we proceed to the analysis of the GC system properties in Section 6, first in terms of the properties of the total GC system and then in terms of the subpopulations and their characteristics. We discuss results in Section 7 before concluding.

\section{Subaru/Suprime Cam Data}
\subsection{Observations and data reduction}
On 2008, April $1^{st}$ we obtained $35 \times27$ arcmin three-filter imaging of NGC 4365 using Subaru/S-Cam. Conditions were good and the worst seeing was $\sim0.8$ arcsec, exposure times were $5\times130$ s, $5\times70$ s and $5\times60$ s for $g'$, $r'$ and $i'$ filters respectively. The images were bias subtracted, flat field corrected and stacked using the SDFRED package \citep{SDFRED1,SDFRED2} and put onto the USNO-B2 astrometric system. Figure \ref{fig:footp} shows a combination of the $g'$, $r'$ and $i'$ filter images. The S-Cam instrument has a pixel scale of $0.202$ arcsec.

The NGC 4365 galaxy light was modelled and subtracted in each of the three filters using the IRAF task ELLIPSE before an object detection algorithm was employed. This was done to increase the success with which the DAOFIND detection algorithm finds faint objects in the central regions of the galaxy (it is unsuccessful at finding objects in areas in which the background surface brightness varies). %The galaxy light was subtracted because the high surface brightness near the centre of the galaxy lessens the probability that the DAOFIND detection algorithm will find faint GCs in that area. 
The standard deviation of the background, after galaxy subtraction, was $\sigma = 8.64$ counts in $g'$, $\sigma = 11.16$ counts in $r'$ and $\sigma = 13.15$ counts in $i'$. Detection thresholds of $2.7\sigma$, $3.0\sigma$ and $2.1\sigma$ in $g'$, $r'$ and $i'$ respectively were used so as to probe as deep as possible in all areas of the images, with the assurance that the selection criteria employed later (see Section 4.2) would remove most spurious detections (due to cosmic rays, background fluctuations and galaxy subtraction artifacts).%These values were input to DAOFIND to detect objects on the images and later to PHOT to determine the errors on magnitudes for those objects. For all images the full width half maximum was 4.0 pixels. Detection threshold at 2.5$\sigma$ and FWHM for finding at 2 pixels. Finding parameters were chosen to ensure that all possible objects were detected and spurious detections were later excluded by a series of selection criteria detailed in Section 3 of this paper.
\subsection{Photometry and calibration}

\begin{figure*}\centering
 \begin{minipage}[t]{0.49\textwidth}\centering
 \includegraphics[width=1.0\textwidth]{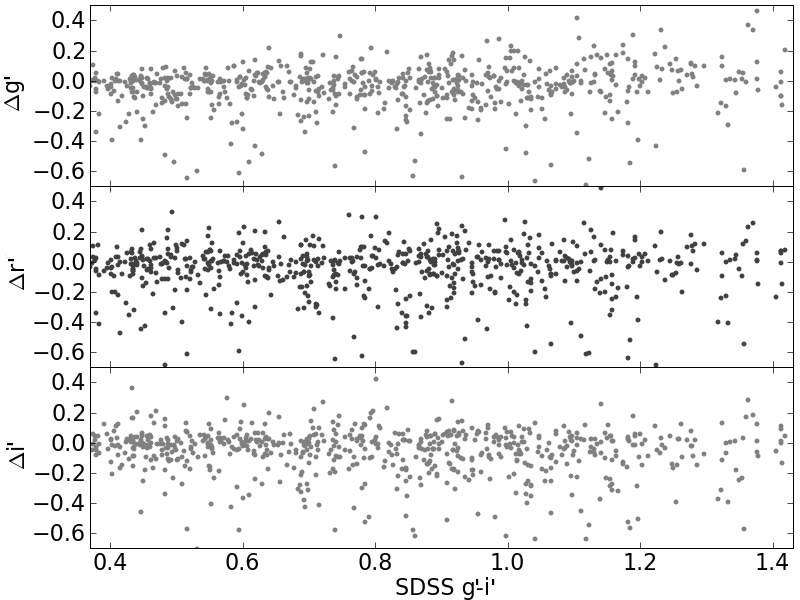}
 \end{minipage}
 \begin{minipage}[t]{0.49\textwidth}\centering
 \includegraphics[width=1.0\textwidth]{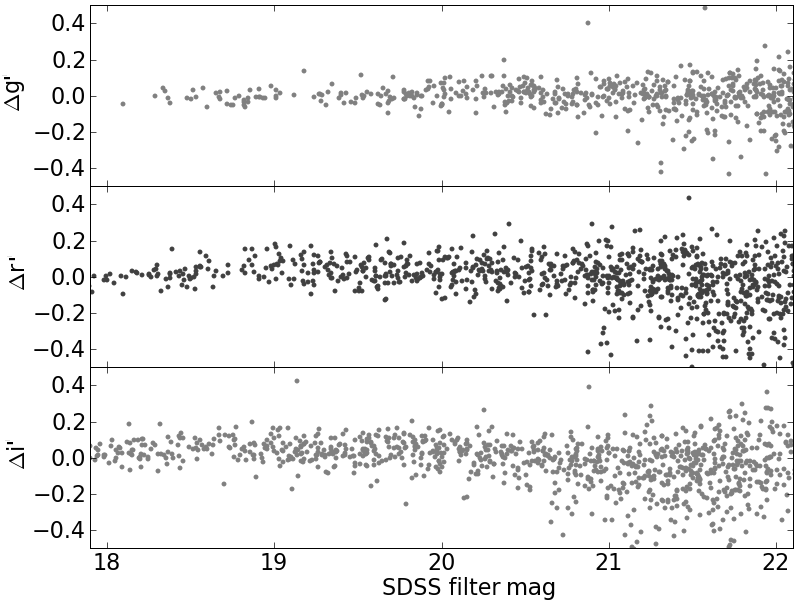}
 \end{minipage}
 \caption{Calibration of S-Cam photometry to SDSS filters. The difference between zeropoint corrected S-Cam and SDSS magnitudes for $g'$, $r'$ and $i'$ filters are plotted from top to bottom. \textbf{Left:} The residuals are plotted against the ($g'-i'$) colour and the colour range is restricted to that expected for GCs. All three filters show asymmetry in their distributions, scattering to brighter S-Cam magnitudes. \textbf{Right:} The residuals are shown against SDSS magnitude.}
 \label{fig:calib}
\end{figure*}

%The derivation of the aperture correction and subsequently the zeropoint for image calibration required point like objects. The method used to determine point sources will be described in the first part of Section 3, as it forms part of a series of selection criteria in determining a GC sample for NGC 4365. 
The point-like objects (see Section 4.2.1 for details on point source determination) brighter than $i' =22$ were used to define an optimum aperture for which to extract the photometry.
%and the aperture correction which needs to be applied to that value to account for the small amount of light beyond the aperture boundary. 
The photometry was extracted at a radius of 3.5 pixels in $g'$ and 3.0 pixels in $r'$ and $i'$ (the seeing was $\sim0.6$ in $r'$ and $i'$ but $\sim0.8$ in the $g'$ filter). The radius of extraction was chosen as a trade-off between increasing error from background noise and increasing uncertainty in the correction for light outside the extraction aperture (i.e.\ an aperture correction). 
%A larger aperture includes more noise but a smaller aperture needs a bigger aperture correction and will result in a bigger error in deriving this value averaged over all objects in the field. 
The aperture correction was $-0.382\pm0.016$ mag in $g'$, $-0.263\pm0.012$ mag in $r'$ and $-0.249\pm0.011$ mag in $i'$. The sky value was calculated as the mode of all pixels in an annulus between 15 and 20 pixels from the centre of the object.% and subtracted from the source in counts.

Standard stars were not observed with NGC 4365, instead the photometry was calibrated using bright point-like objects %(SDSS i magnitude between
($18 <i'< 22$) detected in the NGC 4365 field that were also found in the Sloan Digital Sky Survey (SDSS) catalogue. Both PSF and model derived SDSS magnitudes were used for this and no systematic difference was found in the results. There were 1822, 2174 and 2813 cross matched objects for calibration in $g'$, $r'$ and $i'$ filters respectively and the photometric calibration zeropoints were determined from a best fit linear relation between the SDSS catalogue magnitudes and the S-Cam instrumental magnitudes. These were found to be $zp_{g'} = 27.64\pm0.03$, $zp_{r'} = 27.76\pm0.08$ and $zp_{i'} = 27.72\pm0.10$ on the AB photometric system. The zeropoint corrected residuals of point-like objects in each filter are shown in Figure \ref{fig:calib}, plotted against $g'-i'$ colour (left) and filter magnitude (right). 
The SDSS and S-Cam $g'$, $r'$ and $i'$ filters show good agreement with no obvious systematic colour or magnitude trend. %It was also not necessary to apply an independent correction for airmass effects as the SDSS photometry is airmass independent and airmass would affect the S-Cam field of view similarly everywhere.

It was necessary to correct for foreground dust extinction as such a correction was not applied to the SDSS calibration objects. We used the \citet{Dustmaps} dust maps to calculate the extinction correction in each filter. The values were compared at different positions across the field and not found to vary significantly. For $g'$, $r'$ and $i'$ filters the extinction correction values used were $A_{g'}=0.081$, $A_{r'}=0.060$ and $ A_{i'}=0.045$ mag. %, which are consistent with those quoted by the SDSS catalogue.
Hereafter we quote extinction corrected magnitudes and colours. 
%$$m_{obj}=m_{inst}+ZP+A_{dust} $$
%We show the transformation from instrumental magnitude to the object's apparent magnitude including photometric zeropoint and dust extinction corrections.

\section{HST/ACS Data}
\subsection{Observations and object measurement}
We obtained, from the Hubble Legacy Archive, $g$ and $z$ ($g$ and $i$ for one pointing) filter imaging for eight separate pointings of the HST/ACS instrument around NGC 4365. Table \ref{tab:acsobs} summarises these observations and Figure \ref{fig:footp} shows the ACS pointing footprint on the S-Cam image. These archival data probe down to $z =25.2$, at a $50$ per cent completeness level \citep{VCS12}. The resolution of HST/ACS data (pixel scale is $0.05$ arcsec) partially resolves GCs at the distance of NGC 4365, 23.1 Mpc away \citep{FCS5}. % of the intriguing GC system of NGC 4365.%The central pointing extends out to three arcminutes from the centre of NGC 4365 and was imaged in 2003 by the Virgo Cluster Survey team in the F475W and F850LP filters, almost equivalent to the SDSS g and z filters respectively. Six additional pointings, taken with the same filters, overlap with the central pointing and extend to five and a half arcminutes from the galaxy centre. These pointing were imaged by Sivakoff et al.\ (2007) in 2006 \citep{Fr03} to investigate the connection between GCs and Low-mass X-ray binaries. One pointing was imaged eight and a half arcminutes from the centre of NGC 4365 in late 2003 with an F775W filter (almost equivalent to the SDSS i filter) in place of the F850LP filter, by an independent group to measure cosmic shear.
%In 2003 the Virgo Cluster Survey team obtained imaging in the F475W and F850LP filters of the central three arcminutes of NGC 4365, this data was reduced and published as part of the ACS Virgo Cluster Survey paper series. Later that year an independent group imaged a field on the outskirts of NGC 4365 ($8.5'$ from the galaxy centre) with F475W and F775W filters in order to measure cosmic shear and in 2006 imaging very similar to the original central ACS Virgo Cluster Survey observations was taken in 6 fields surrounding and partly overlapping the central three arcminutes to extend the ACS imaging of NGC 4365 out to five and a half arcminutes. These observations were taken to investigate the connection between GCs and Low-mass X-ray binaries, published with data on other Virgo Cluster galaxies in Sivakoff et.\ al.\ 2007.
\begin{table}
 \centering
  \caption{Imaging obtained from the Hubble Legacy Archive.}
 \begin{tabular}{| c | c c | c | c | c| }
 \hline
  &\multicolumn{2}{|c|}{\textbf{Central}} &\multicolumn{2}{|c|}{\textbf{Exposures}} & \textbf{HST} \\
 & \textbf{R.A.} & \textbf{Dec.} & \textbf{Filter} & \textbf{Time} (s) & \textbf{ID} \\
 \hline \hline
  &  &  & F475W & 750 & \\
 1 & 12:24:27.0 & 07:19:20.8 & F850LP & 1210 & 9401 \\ \hline
  &  &  & F475W & 680 & \\
 2 & 12:24:17.5 & 07:21:09.5 & F850LP & 1170 & 10582 \\ \hline
  &  &  & F475W & 680 & \\
 3 & 12:24:16.0 & 07:17:37.1 & F850LP & 1170 & 10582 \\ \hline
  &  &  & F475W & 680 & \\
 4 & 12:24:26.3	& 07:15:44.8 & F850LP & 1170 & 10582 \\ \hline
  &  &  & F475W & 680 & \\
 5 & 12:24:36.4	& 07:17:25.1 & F850LP & 1170 & 10582 \\ \hline
  &  &  & F475W & 680 & \\
 6 & 12:24:41.3 & 07:20:18.2 & F850LP & 1170 & 10582 \\ \hline
  &  &  & F475W & 680 & \\
 7 & 12:24:31.5	& 07:22:20.4 & F850LP & 1170 & 10582 \\ \hline
  &  &  & F475W & 1744 & \\
 8 & 12:24:07.2 & 07:12:11.3 & F775W & 1624 & 9488 \\
 \hline
 \end{tabular}
 \label{tab:acsobs}
\end{table}

%\begin{figure}
% %\begin{minipage}[t]{0.49\textwidth}\centering
%  \includegraphics[width=0.45\textwidth]{magcheck2.png}
%  \caption{The difference between previously published magnitudes and those derived in this work for objects in the central HST/ACS field. Results from filters F475W (plotted in blue) and F850LP (plotted in red) are compared and both show good agreement with previous work. Measurements agree to within $\sim 0.1$ mag even for relatively faint objects and no systematic offset is seen.}
%  \label{fig:mag}
% %\end{minipage}
%\end{figure}
% %\hspace{1mm}
%\begin{figure}
% %\begin{minipage}[t]{0.49\textwidth}\centering
%  \includegraphics[width=0.45\textwidth]{radcheck.png}
%  \caption{The difference between previously published sizes and those derived in this work for objects in the central HST/ACS field. Symbols as in Figure \ref{fig:mag}. Both filters show a slight systematic offset in size measured with half light radii measured in this work $\sim 0.005''$ smaller. This is attributed to the different methods for determining radius between this work and previously published results.}
%  \label{fig:rad}
% %\end{minipage}
%\end{figure}

%\begin{figure}[*h]\centering
% \includegraphics[width=0.8\textwidth]{magcheck2.png}
% \caption{The difference between magnitudes }
%\end{figure}
%
%\begin{figure}[*h]\centering
% \includegraphics[width=0.8\textwidth]{radcheck.png}
% \caption{Radii measurements}
%\end{figure}

The eight individual ACS fields were analysed using a custom built pipeline (see e.g.\ \citealt{St06} and \citealt{Sp06}) to find small, round objects and measure their magnitudes and half light radii (henceforward referred to as object size). For details on the methods used by the pipeline including point spread function determination see \citet{St06}.%as well as sharpness, roundness and major to minor axis ratio in each band. 
The object lists from each field were collated, with the arithmetic mean taken of the sizes and magnitudes in the field overlap areas, and associated errors adjusted to reflect the more accurate measurement.

The measured magnitudes and sizes are compared with those published for objects in the central field \citep{VCS16}. The magnitude measurements show no evidence of a statistically-significant systematic offset from the published data. The scatter increases to fainter objects but is on the order of 0.1 magnitudes. The size measurements in this work show a small $\sim 0.005$ arcsec systematic offset to smaller sizes attributed to differences in size measurement techniques.

\section{GC candidate Selection}
%The information that can be obtained from three colour photometry on a ground based, seeing limited instrument can be used to determine a statistically accurate sample of the objects in NGC 4365's GC population. By using a combination of various selection criteria a spatially complete, magnitude limited catalogue of GCs in NGC 4365 can be obtained from the Subaru/S-Cam imaging. GC samples derived for other galaxies with similar methods have been shown, by comparison with subsequent spectroscopic observations, to have contamination rates as low as 1-2\%.
\subsection{HST/ACS GCs}

\begin{figure}\centering
 \includegraphics[width=0.47\textwidth]{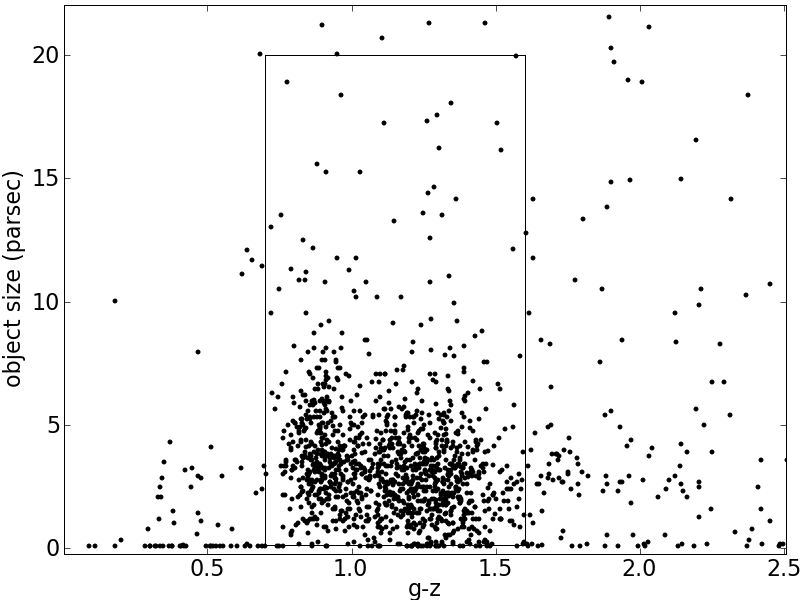}
 \caption{Size in parsec vs.\ $g - z$ colour distribution for GC candidates brighter than the turnover magnitude of $z=23.4$, detected in the HST/ACS fields. The GC candidate selection criteria is marked with a box. The blue GC candidates tend to have larger average sizes and the red GC candidates have a larger dispersion in colour.} %Size was derived using the ACS pixel scale of $0.05''$ per pixel.
 \label{fig:colrad}
\end{figure}

\begin{figure}\centering
 \includegraphics[width=0.47\textwidth]{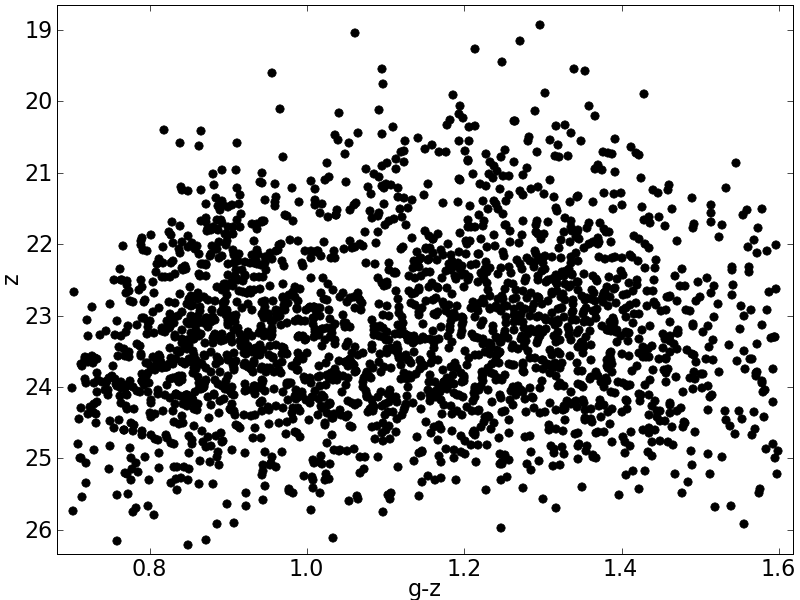}
 \caption{Colour magnitude diagram for GC candidates selected from HST/ACS imaging.}
 \label{fig:cmda}
\end{figure}
GC candidates were selected from the HST/ACS images based on their size, colour and magnitude. The GC candidate distribution is selected to have colours $0.7<g - z<1.6$ and sizes $0.1\,\mathrm{pc}<\mathrm{r}_h<20\,\mathrm{pc}$. The choice of $g - z$ colour upper and lower bounds was based on the colour size diagram shown in Figure \ref{fig:colrad}. There is a clear drop off in the density of objects bluewards of 0.75 and a similar but less clear drop off redwards of 1.6 ($-2.4<\mathrm{[Fe/H]}<0.18$ using the empirical transformation of \citet{VCS9}). % Fe/H = -6.21+(5.14\pm0.67)(g-z) where 0\le g-z\le 1.05 and Fe/H = -2.75+(1.83\pm0.23)(g-z) where 1.05<g-z\le 1.45
%Not all GCs in the GC system of NGC 4365 are resolved by HST/ACS and particularly some of the red GCs become small enough in size to be indistinguishable from stellar point sources. 
While most of the GC candidates in Figure \ref{fig:colrad} are resolved, the redder GC candidates show a size distribution that overlaps with objects of zero size (objects indistinguishable from stellar point sources). A size cut at 0.1 pc ($10^{-4}$ arcsec) was made because at smaller sizes the objects did not show a distribution that was centrally concentrated on the galaxy 
%or clustering in colour space 
and therefore does not sample mainly GCs. A generous upper bound on size of $20\,\mathrm{pc}$ ($0.2$ arcsec) was employed to exclude background galaxies but include possible Ultra Compact Dwarfs (UCDs) in the system. Lastly, only objects fainter than $z =19$ (M$_z=-12.8$ mag) are considered GC candidates of NGC 4365, following the convention of the ACS VCS published catalogue \citep{VCS16}. The catalogue contains no objects fainter than $z = 26.2$ mag. Objects brighter than $z=19$ are most likely to be stars. % and only very few of these objects might be Ultra Compact Dwarfs. %The sample of NGC 4365 GCs determined in this way will exclude a few GCs and include a few objects that are not GCs but traces the properties of NGC 4365 GC system when considered as a statistical sample.
\subsection{Subaru/SuprimeCam GCs}
\subsubsection{Point source determination}
\begin{figure}\centering
 \includegraphics[width=0.47\textwidth]{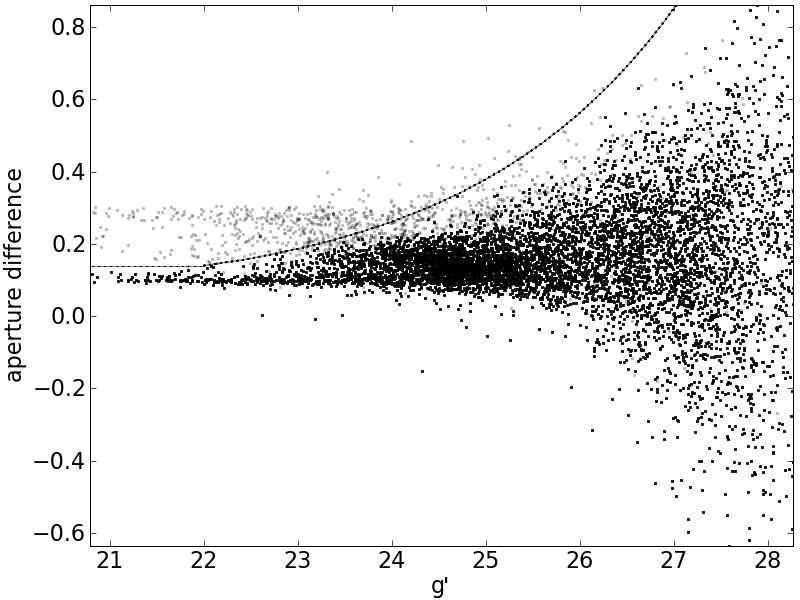}
 \caption{The difference in flux between two aperture radii (the smaller of which is the extraction aperture) vs.\ aperture corrected magnitude at the extraction aperture for all objects found by the DAOFIND detection algorithm. Objects determined to be point-like in all three S-Cam filters are shown in black dots, overlapping the total object detection in the $g'$ filter (grey dots). The dotted line shows the cutoff between point-like objects and extended objects in the $g'$ filter. Bright point-like objects have aperture differences of $\sim0.1$ magnitudes but fainter objects show significant scatter in measured magnitudes. A difference in magnitude between the inner and outer aperture of up to $\sim0.6$ mag is required to include fainter point-like objects.}
 \label{fig:ptsel}
\end{figure}
At the distance of NGC 4365, GCs are not resolved by our Subaru observations and consequently they cannot be separated from stars in our Galaxy via a size distinction. By selecting only objects that are unresolved in the S-Cam imaging background galaxies are excluded from the analysis of NGC 4365's GC system. Here the distinction between a point-like object (mostly stars and NGC 4365 GCs) and an extended object (galaxies) is determined by a measure of the flux difference between two aperture radii. Shown in Figure \ref{fig:ptsel} are the objects that were identified to be point-like in all three filters. Magnitudes were extracted for two apertures different in radius by a half or full pixel (filter dependent) both centered on the object. Extended objects have extra light in the larger aperture compared to point-like objects. %In this way point-like objects can be distinguished from extended objects. Using the $g'$ filter as an example, objects brighter than $g' \sim22$ are considered point sources if the aperture difference is $\le 0.15$ but for fainter objects (where photometric uncertainty becomes important) the distinction between point-like objects and extended objects increases to fainter magnitudes following an exponential function.
%Objects with a large difference in flux between an aperture with a reasonably small radius and another with a radius larger by a pixel or half a pixel have extra light beyond what is expected for point sources and are taken to be extended objects. At this stage it was still considered prudent to be inclusive rather than exclusive and on each image the cutoff between point sources and extended sources was placed to ensure that objects with ambiguous light profiles, upon visual inspection, were regarded as point sources. This was only a consideration for faint objects. 

The lists of point-like objects for each filter were cross-matched in position and only objects classed as point-like in all three filters were kept in the sample. %Before colour selection criteria were applied to the objects found in the images the sample of possible objects was reduced by excluding all objects that did not have a counterpart in all three bands, matched to $3.5^{\circ}\times10^{-4}$ or $1.26''$ in the world coordinate system. 
This was done using the IRAF task TMATCH with a tolerance in positional offset between images of $1.26$ arcsec (this was determined by the uncertainty in astrometry at the image edges). Because the photometry is deeper in the $i'$ filter than either $r'$ or $g'$ filters this procedure likely excluded genuine faint GCs from the analysis that were detected in $i'$ but not in either of the other filters. 
%The likelihood that an object detected in one or two of the bands but not in a third is a spurious detection is however much higher than the likelihood that such an object was not detected in the other two bands due to photometric incompleteness and furthermore 
Using a cross-matching technique spurious detections (due to cosmic rays etc.) in any of the three filters were eliminated and only objects with photometry in all three filters selected. %Three filter photometry is needed for the $(g-r)_0$ vs.\ $(r-i)_0$ colour-colour selection to determine a GC candidate list with very low contamination from background galaxies or foreground stars.
\subsubsection{Colour-colour and magnitude selection}

\begin{figure}\centering
 \includegraphics[width=0.47\textwidth]{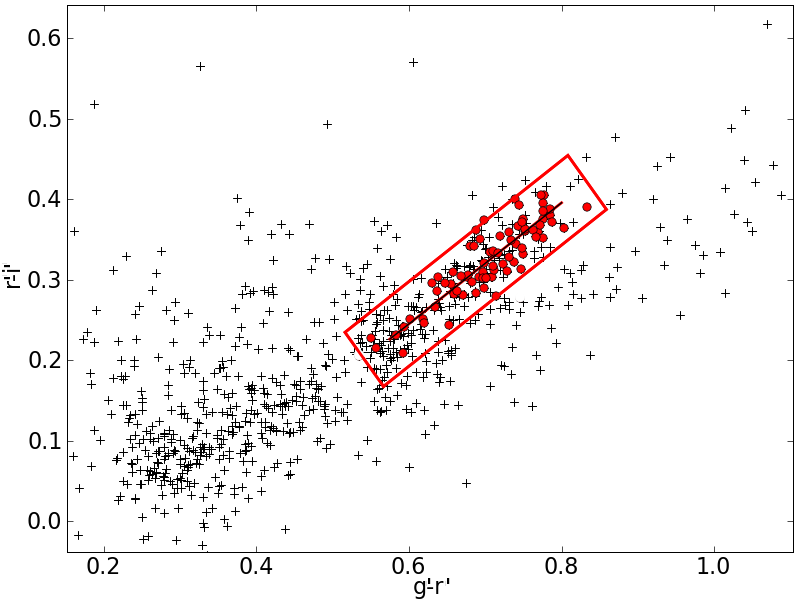}
 \caption{Colour selection of GCs in the S-Cam field-of-view based on the $r' - i'$ vs.\ $g' - r'$ colour of matched objects in the HST/ACS data. The objects shown are all brighter than $i' =22$ mag and have magnitude errors smaller than 0.02 for visual clarity. S-Cam detected point-like objects are plotted in black plusses and HST/ACS matched objects are overplotted as red dots. The HST/ACS matched objects define a very tight sequence (black line) that is used to define the area from which S-Cam objects are selected as GC candidates (red box).}
 \label{fig:colsel}
\end{figure}

\begin{figure}\centering
 \includegraphics[width=0.47\textwidth]{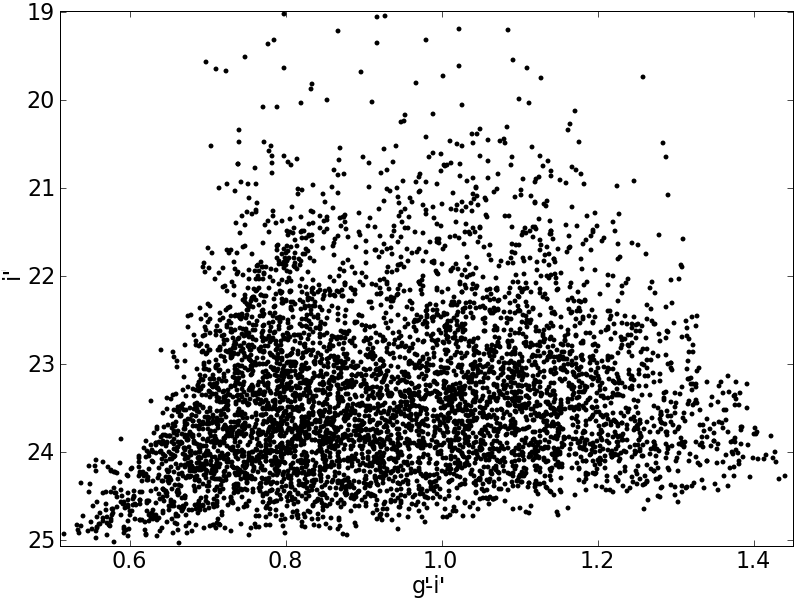}
 \caption{Colour magnitude diagram for GC candidates selected from S-Cam imaging.}
 \label{fig:cmds}
\end{figure}
As shown in Figure \ref{fig:colrad} the HST/ACS is able to partially resolve GCs at the distance of NGC 4365 and can therefore directly distinguish between unresolved foreground stars, GCs and resolved background galaxies. We used the central HST/ACS pointing to determine where on a S-Cam $r' - i'$ vs.\ $g' - r'$ diagram the GC sequence lies by matching S-Cam point source detections with ACS GC candidates. A line was fitted to the brightest S-Cam objects found to be GC candidates in the ACS field and used to make a box $2\sigma$ in each direction from the best fit GC sequence line. Objects were considered to be consistent with the GC colour-colour definition if the standard error in their photometry placed them within the box (see Figure \ref{fig:colsel}). The selected distribution contains objects with colours $0.6<g-i<1.4$, corresponding to metallicities of $-2.1<\mathrm{[Fe/H]}<0.7$ when the \citet*{Le10b} empirical transformation is used.%Only objects with photometric errors smaller than 0.1 mag in all bands were considered GC candidates.

An upper limit on the magnitude of any GC candidate was set to $i' =19$ ($M_{i'}=-12.8$ mag) by the brightest GC found by HST/ACS in the central field of NGC 4365. %The magnitude of this object was i $=19.02$ in the S-Cam data, therefore the upper limit on the magnitude for objects in NGC 4365's GC system was set to i $=19$. 
A limit of 0.1 mag in photometric error resulted in a lower limit of $i' \approx25$ mag on the blue end of the colour-colour selection and $i' \approx24.5$ mag on the red end of the colour-colour selection.
\subsection{Contamination and completeness}
At galactocentric radii smaller than $0.9$ arcmin S-Cam detection numbers drop due to galaxy subtraction artifacts and at radii greater than $3.4$ arcmin ACS detections are no longer spatially complete due to the tiling pattern of HST/ACS pointings. Therefore only in an annulus between $0.9$ and $3.4$ arcmin from the centre of NGC 4365 is it possible to compare the object detections and photometry for the S-Cam and ACS imaging. This was done to determine the level of contamination in the S-Cam GC sample (contamination is defined as non GC objects that meet the colour, size and magnitude conditions set for the S-Cam photometry) as well as to determine the photometric completeness of our S-Cam imaging. This analysis is relative to the HST/ACS sample but the ACS imaging is almost perfectly complete several magnitudes deeper than our Subaru/S-Cam imaging \citep{VCS16}. 

\begin{figure}\centering
 \includegraphics[width=0.47\textwidth]{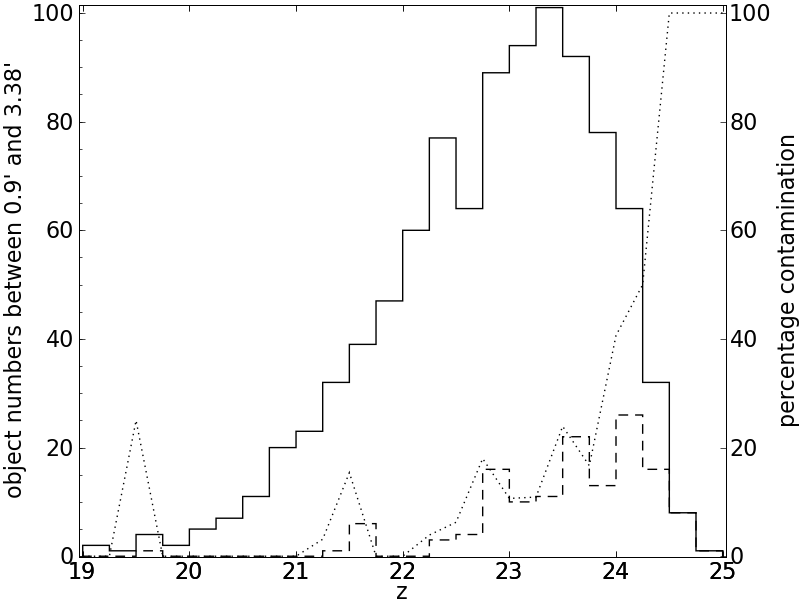}
  \caption{Estimate of contamination in the S-Cam data by comparison with HST/ACS matched objects. The solid line shows the number of S-Cam GCs in each magnitude bin, the dashed line shows the number of those objects that are determined to be contaminants by HST/ACS imaging and the dotted line shows the percentage contamination of the S-Cam GC sample as a function of magnitude.}
  \label{fig:contam}
\end{figure}

The contamination percentage of the sample in S-Cam was determined by keeping record of the GC candidates detected in the ACS imaging but removed by the conditions outlined in Section 4.1. %The uncatalogued objects were visually inspected and all determined to be faint and diffuse. 
At a magnitude of $z =24.2$, 50 per cent of S-Cam GC candidates are determined to be contamination and at $z =24.5$ mag essentially all of the S-Cam GC candidates are contaminants. Across the whole GC candidate sample the contamination determined in this way is $4.14 \pm 0.35$ objects $\mathrm{arcmin}^{-2}$ and if the sample is restricted to only objects brighter than $z =23.4$ mag this value drops to $1.23 \pm 0.19$ objects $\mathrm{arcmin}^{-2}$. The luminosity function of GCs can be well described by a Gaussian distribution peaked at the turnover magnitude of $M_z\approx -8.4$ mag \citep{VCS12}, which corresponds to $z =23.4$ mag at the distance of NGC 4365 and all GC candidates described in further analysis are brighter than this turnover magnitude.

\begin{figure}\centering
 \includegraphics[width=0.47\textwidth]{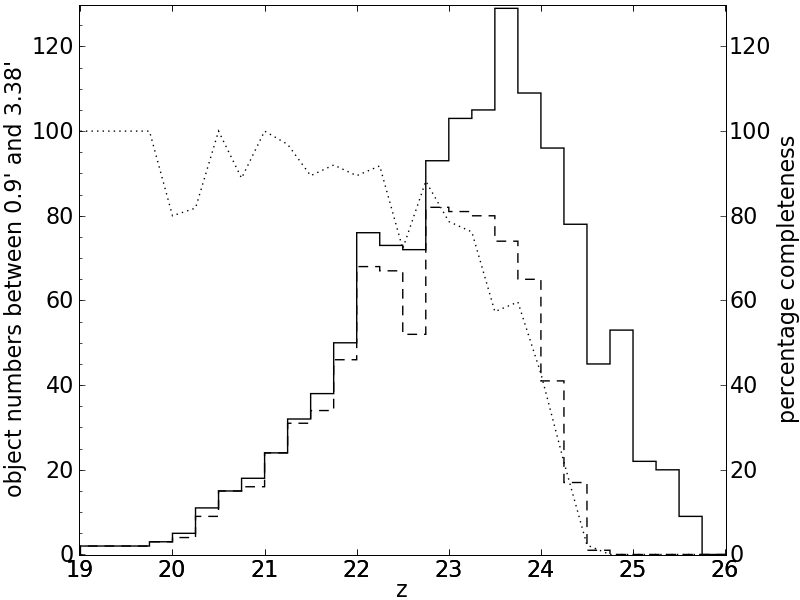}
  \caption{Estimate of the completeness of the S-Cam photometric sample by comparison with HST/ACS photometry. The solid line shows the number of HST/ACS GCs in each magnitude bin, the dashed line shows the number of those objects that were also detected in the S-Cam imaging and the dotted line shows the percentage completeness of the S-Cam photometry as a function of magnitude.}
  \label{fig:compl}
\end{figure}

The completeness of the S-Cam imaging was determined by keeping record of how many of the HST/ACS detected objects were also detected by the Subaru/S-Cam imaging at different magnitudes. The S-Cam imaging is found to be 50 per cent complete at $i'=23.8$ mag and 70 per cent complete at $i=23.6$ mag as shown in Figure \ref{fig:compl}.

%\begin{table}
% \centering
%  \caption{Completeness in the central regions of the ACS imaging.}
% \begin{tabular}{c c c}
% \hline
% \textbf{Galactocentric} & \textbf{Completeness} \\			% & Galaxy Surface
% \textbf{Radius} & \textbf{Fraction} \\								% & Brightness
% \hline \hline
%%$0.1' $& 6.23 count/pix$^2$ & 0.325 \\
%$0.2$ arcmin & 0.933 \\				%  & 1.53 count/pix$^2$
%$0.325$ arcmin & 0.989 \\			%  & 0.735 count/pix$^2$
% \hline
% \end{tabular}
% \label{tab:acscompl}
% \end{table}

In the very central regions (radii $<0.325$ arcmin) of NGC 4365 the HST/ACS detection of GC candidates is inhibited by the high surface brightness of the galaxy. In \citet{VCS12} the completeness of objects is tabulated as a function of object size, object magnitude and surface brightness of the galaxy at the location of the GC candidate. For this analysis it is only necessary to consider completeness as a function of radius. We measured the average surface brightness of the galaxy in $0.1$ arcmin annuli on the ACS imaging. We compared that to the completeness fraction for an object of mean size $=0.039$ arcsec ($4.3$ parsec) and $z=23.4$ mag, which is representative of the GC candidate sample. At distances of $0.2$ and $0.325$ arcmin from the galaxy centre the ACS imaging is 93.3 and 98.9 per cent complete.

%\begin{figure}[ht]
% \begin{minipage}[t]{0.5\textwidth}\centering
%  \includegraphics[width=1.0\textwidth]{contam_sbr.png}
%  \caption{Estimate of contamination of NGC 4365's GC system determined from the S-Cam imaging by comparison with HST/ACS matched objects. The solid line shows the number of S-Cam GCs in each magnitude bin, the dashed line shows the number of those objects that are determined to be contaminants by HST/ACS imaging and the dotted line shows the percentage contamination of the S-Cam GC sample as a function of magnitude.}
% \end{minipage}
% \hspace{1mm}
% \begin{minipage}[t]{0.5\textwidth}\centering
%  \includegraphics[width=1.0\textwidth]{complete_sbr.png}
%  \caption{Estimate of the completeness of the S-Cam photometric sample by comparison with HST/ACS photometry. The solid line shows the number of HST/ACS GCs in each magnitude bin, the dashed line shows the number of those objects that were also detected in the S-Cam imaging and the dotted line shows the percentage completeness of the S-Cam photometry as a function of magnitude.}
% \end{minipage}
%\end{figure}

\subsection{Overview of GC selection}
% LEE SUGGESTS LEAVING IT OUT
The selection of GC candidates from the Subaru/S-Cam photometry was done based on object size, magnitude and locus in colour-colour space compared to GC candidates obtained from HST/ACS photometry. Point-like objects were selected in each S-Cam filter using a limit on the flux difference between two radii apertures (see Section 4.2.1). The astrometry of the ACS GC catalogue was shifted to that of the S-Cam catalogue. The positions of objects classed as point-like in all S-Cam filters were compared with the positions of GC candidates obtained from the ACS photometry (see Section 4.1 for ACS selection criteria; colours $0.7<g - z<1.6$, sizes $0.1\,\mathrm{pc}<\mathrm{r}_h<20\,\mathrm{pc}$ and magnitudes $26.2\leq z \leq19.0$). The matched S-Cam point-like objects were used to determine the locus of GCs on a $r' - i'$ vs.\ $g' - r'$ diagram. %This distribution was characterised by a straight line of finite length and defined standard deviation, all objects that had colours and associated errors that were consistent with a $2\sigma$ box around the line, as well as magnitudes brighter than $i = 19.0$ and photometric errors $<0.1$, were considered GC candidates (see Section 4.2.2). 
No contaminating objects could be visually identified on inspection of the GC candidates brighter than $i =21$ mag. Finally, estimates of the contamination and completeness of the S-Cam catalogue were determined by comparison with the ACS catalogue (see Section 4.3). The full catalogue of NGC 4365's GC candidates, found either ACS or S-Cam imaging, is included as online supplementary material in \textit{FullGlobList.dat}.

\section{Analysis of the galaxy light}
\subsection{Parameters of the isophotal fits}
\begin{figure*}\centering
 \includegraphics[width=0.95\textwidth]{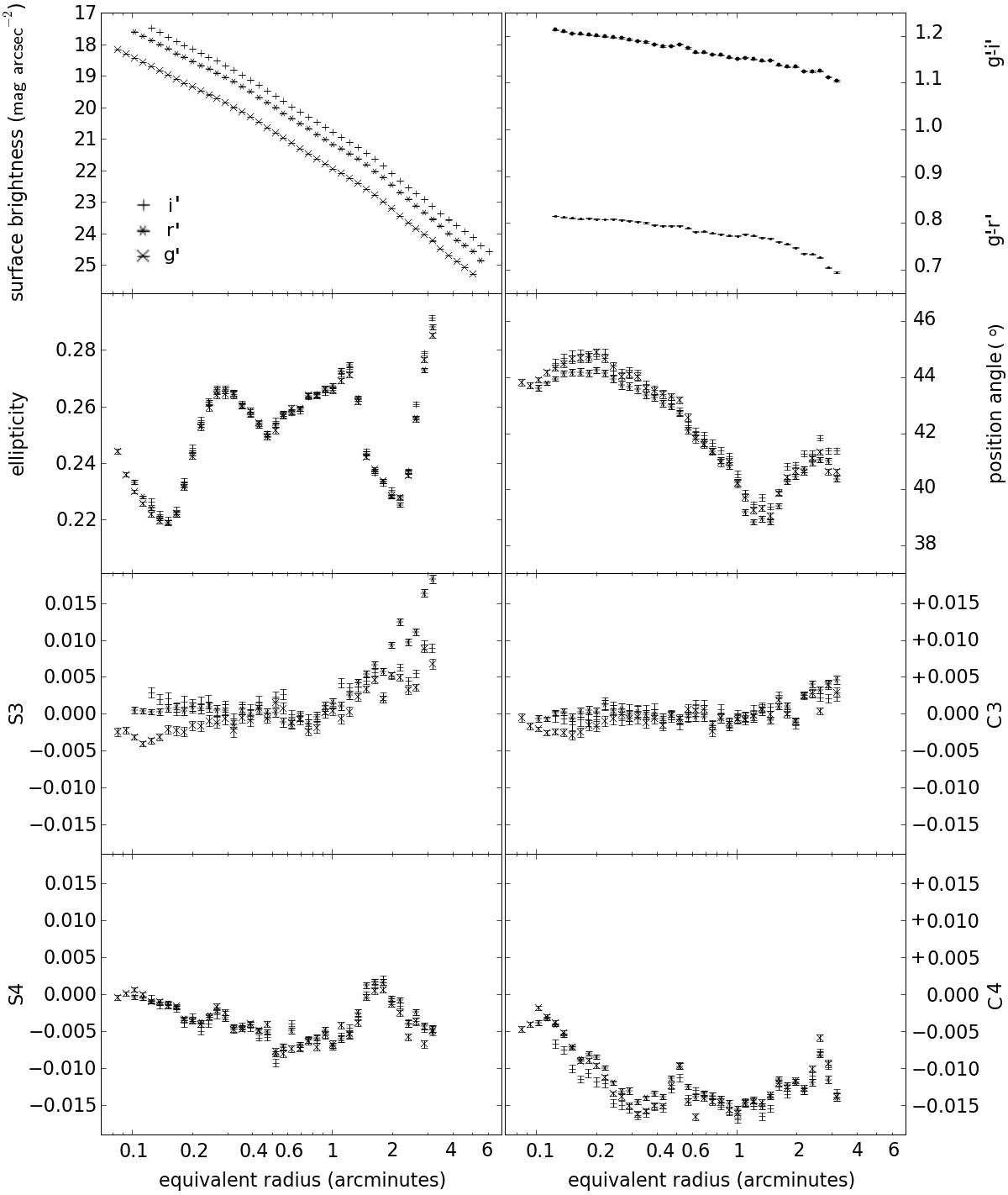}
 \caption{Isophotal parameters for $g', r'$ and $i'$ filters from Subaru/S-cam photometry of NGC 4365. From left to right and top to bottom the parameters plotted are: surface brightness of the galaxy light in $g'$ (crosses), $r'$ (stars) and $i'$ (plusses) filters; galaxy colour; ellipticity of the fitted isophotes in all three filters as labelled before; position angle of the fitted isophotes in all three filters as labelled before; higher order Fourier coefficients to the sine (S3 and S4) and cosine (C3 and C4) terms. All parameters are plotted against equivalent galactocentric radius, calculated as the geometric mean of the semi-major and semi-minor axis lengths of the fitted ellipse. The fitted values become unstable beyond $3$ arcmin. Isophotal parameters for the i' filter are included as online supplementary material in \textit{GalLightTable.dat}.}
 \label{fig:galprop}
\end{figure*}

We used the IRAF task ELLIPSE to model and subtract the galaxy light from the Subaru/S-Cam photometry. From the model of the galaxy light we obtain the surface brightness, ellipticity, position angle and higher order Fourier terms (S3, S4, C3, C4) in $g', r'$ and $i'$ filters for NGC 4365. Results are plotted in Figure \ref{fig:galprop} and briefly discussed here. We have surface brightness information for all three filters between $\sim0.1$ and $5$ arcmin. We fit S\'{e}rsic profiles \citep{Gr05}, i.e.\
\begin{equation}
\mu(R)=\mu_e + \frac{2.5b_n}{ln 10}\left[\left(\frac{R}{R_e}\right)^{\frac{1}{n}}-1\right]
\end{equation}
to the three surface brightness profiles, where $R_e$ is the effective radius of the galaxy, $\mu_e$ is the surface brightness at that radius, $n$ is the shape parameter of the S\'{e}rsic profile and $b_n=1.9992n-0.3271$. The fitted values are tabulated in Table \ref{tab:galSers}.

\begin{table}\centering
 \begin{tabular}{cccc}
 \hline
 \textbf{Filter} & {\boldmath$\mu_{e}$}  & $\mathbf{n}$  & $\mathbf{R_{e}}$ \\
 & (mag $\mathrm{ arcsec}^{-2}$) & & (arcmin) \\ \hline \hline
 g & $21.87\pm0.11$ & $6.02\pm0.21$ & $2.40\pm0.14$ \\
 r & $21.46\pm0.12$ & $5.92\pm0.23$ & $2.06\pm0.12$ \\
 i & $21.20\pm0.11$ & $5.97\pm0.24$ & $2.10\pm0.11$ \\ \hline
 \end{tabular}
 \caption{Values for $g$, $r$ and $i$ S\'{e}rsic profiles fitted to galaxy light.}
 \label{tab:galSers}
\end{table}

The quoted value for NGC 4365's effective radius \citep{Go94,Be92} is $1.1$ arcmin based on a de Vaucouleurs profile (equivalent to a S\'{e}rsic profile with the $n$ parameter set to 4) fitted by \citet{Bu87}. When a de Vaucouleurs profile is fit to the data in this work the effective radius is found to be $1.32\pm0.03$ arcmin but we find this to be a significantly poorer fit than a general S\'{e}rsic profile. For all further analysis we use our $R_e$ value of $2.1\pm0.1$ arcmin ($14.1\pm0.7$ kpc), with an $n$ value of $6.0\pm0.2$. This is close to the value of $1.6\pm0.1$ arcmin ($10.9\pm0.7$ kpc) with $n=5.8$ in $z$ that \citet{Ch10} found and slightly smaller than the value of $3.07\pm0.22$ arcmin ($20.6\pm1.5$ kpc) with $n=7.1\pm0.4$ in $B$ that \citet{Ko09} found.

Results for the ellipticity, position angle and higher order Fourier terms in this work agree with the values \citet{Go94} found in the $B, V$ and $I$ filters and extend $2$ arcmin further, to $\sim1.5 R_e$. The ellipticity measured in this work varies between 0.22 at $0.15$ arcmin and 0.28 at $2.5$ arcmin and is consistent with that seen by \citet{Go94} interior to $1$ arcmin. %However, the measured value for position angle doesn't continue to decrease below $38^{\circ}$ beyond $\sim 1'$ but increases to $42^{\circ}$ between $1'$ and $2.5'$. 
The position angle shows a small twist of $\leq 5^{\circ}$beyond the radial range explored by \citet{Go94}. The S3 parameter (coefficient to the sine term in the fit) shows an indication of increase at the edge of their radial range ($1$ arcmin) which is confirmed in this work (between $1$ and $2$ arcmin) and correspondingly the S4 parameter increases between $1$ and $2$ arcmin. There is no significant difference in C3 and C4 parameters between \citet{Go94} and this work. The C4 disky/boxy parameter remains at values between -0.01 and -0.02 out to $2$ arcmin, indicating that NGC 4365 has boxy isophotal structure to large radii. The galaxy colour becomes bluer with increasing galactocentric radius, changing by $\sim0.1$ mag from $\sim0.1$ to $2.5$ arcmin for both $g-i$ and $g-r$ colour indices. A similar trend, implying a negative metallicity gradient, is seen in the \citet{Go94} B--I colour index.

\section{Analysis of the GC System}
\subsection{Characterising the total GC System}
\subsubsection{Surface Density}
\begin{figure}\centering
 \includegraphics[width=0.47\textwidth]{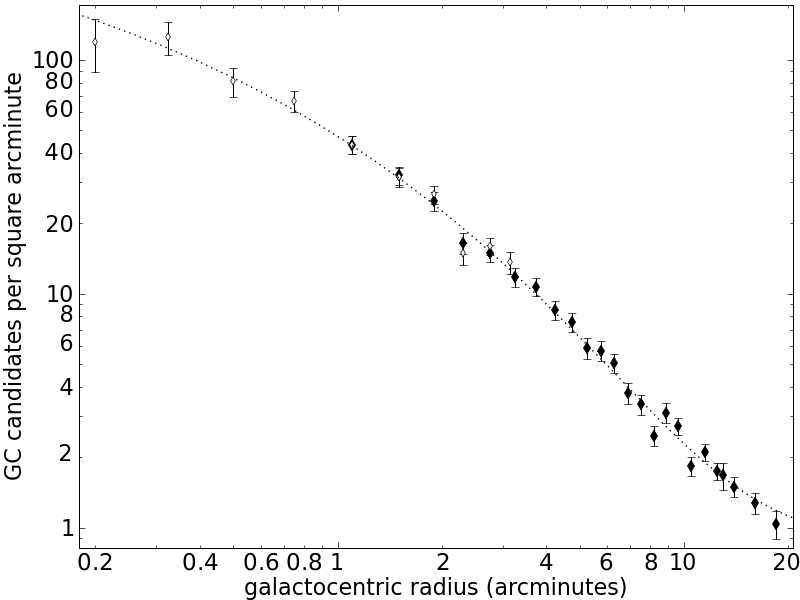}
 \caption{Surface density of GC candidates brighter than the turnover magnitude plotted against galactocentric radius. The S-Cam points (solid points) were calculated using objects brighter than $i'=23.6$ and the ACS points (hollow diamonds) were calculated using objects brighter than $z=23.4$. The innermost two ACS values were corrected for incompleteness using the values discussed in the text. %but the corrected value displayed at $0.1'$ is based on only 1 detection and therefore is unsurprisingly very uncertain. 
 No other normalisation correction has been made to either ACS or S-Cam radial surface density and the remarkable agreement in the region of overlap (between $1$ and $3$ arcmin) is an indication that both samples are almost perfectly complete to the turnover magnitude and contain very little contamination. The line plotted is a fitted S\'{e}rsic profile with a constant background value. See text for further details.}
 %( $I_0 \exp\left[ -b_n \lbrace (\frac{R}{R_e})^{\frac{1}{n}}-1 \rbrace \right]+bg$ ) fit to the data, where $I_0=3.9\pm1.3\,\mathrm{arcmin}^{-2}$, $R_e=6.1'\pm 1.2'$, $n=2.68\pm 0.41$, $bg=0.89\pm 0.13\,\mathrm{arcmin}^{-2}$ and $b_n=1.9992n-0.3271$.
 \label{fig:bgser}
\end{figure}

%\begin{figure}[!ht]\centering
% \includegraphics[width=0.6\textwidth]{Sersic.png}
% \caption{The density points here are derived as described in Figure \ref{fig:bgser} but the background contamination estimate derived in Section 4.3 ($1.2\pm 0.6$objects per square arcminute) has been subtracted before a S\'{e}rsic profile (dashed line) was fitted to the data. The resulting fit values are $R_e=5.04'\pm 0.57$ and $n=2.4'\pm 0.2$.}
% \label{fig:ser}
%\end{figure}
 
By combining the wide field imaging from S-Cam with the spatially resolved imaging from ACS and further restricting the GC candidate sample to objects brighter than the turnover magnitude we can derive a very accurate radial surface density profile over a factor of 100 in radius. To derive surface density profiles, the number of GC candidates in each radial bin were divided by the total  area in that radial bin. The results for both the S-Cam and ACS GC candidate samples are plotted on Figure \ref{fig:bgser}.

%It has been difficult in the past to derive the parameters of profile fits to GC surface density data because there simply haven't been sufficient GC numbers or spatial coverage. Thus historically only the slope of a power-law approximation to the density profile has been quoted.  Here we use only the GC candidates brighter than the turnover magnitude to constrain the profile of the system. 
We expect very little contamination in the density profile, with more than 90 per cent completeness in this magnitude range (see Section 4.3). %if the whole system were considered using data of similar quality to $z\approx 28$ the shape of the profile would be essentially unchanged, only the values of object density would double. 
A S\'{e}rsic profile plus a background parameter, i.e.\ 
\begin{equation}
P(R)=P_e \exp\left( -b_n \left[ \left(\frac{R}{R_e}\right)^{\frac{1}{n}}-1 \right] \right)+bg
\end{equation}
where 
\begin{equation}
b_n=1.9992n-0.3271
\end{equation}
is fitted to the combined S-Cam and ACS radial surface density of GC candidates and shown in Figure \ref{fig:bgser}. $R_e$ is the effective radius of the GC system, $P_e$ is the density at the effective radius and $n$ is the shape parameter of the S\'{e}rsic profile. The fitted parameters are recorded in Table \ref{tab:gcSers}.
%Both methods of profile fitting agree on the effective radius ($5.41'\pm 0.94'$) and n value ($2.5\pm 0.35$) and n value of the S\'{e}rsic profile fit within errors. 
The value for contamination expected from the analysis in Section 4.3 ($1.23\pm 0.19\,\mathrm{arcmin}^{-2}$) agrees within $2\sigma$ with the value from the S\'{e}rsic profile fit ($0.89\pm 0.13\,\mathrm{arcmin}^{-2}$).

\begin{table}\centering
 \begin{tabular}{cccc}
 \hline
 {\boldmath$P_{e}$}  & $\mathbf{n}$  & $\mathbf{R_{e}}$ & $\mathbf{bg}$\\
 ($\mathrm{arcmin}^{-2}$) & & (arcmin) & ($\mathrm{arcmin}^{-2}$) \\ \hline \hline
 $3.9\pm1.3$ & $2.68\pm0.41$ & $6.1\pm1.2$ & $0.89\pm0.13$ \\ \hline
 \end{tabular}
 \caption{Fitted values of the S\'{e}rsic fit to the GC surface density.}
 \label{tab:gcSers}
\end{table}

We are not able to define the edge of NGC 4365's GC system with certainty as the profile shown in Figure \ref{fig:bgser} is still decreasing at $\sim 20$ arcmin ($134$ kpc), which is the very edge of the spatial coverage available to us with S-Cam imaging. In order to confirm the edge of the GC system even larger field of view imaging would be required. We can make an estimate of the total number of GCs associated with NGC 4365 by simply doubling the number of GC candidates brighter than the turnover magnitude ($i'=23.6$ and $z=23.4$). This assumes that there is no inherent asymmetry in the GC luminosity function, the GC system of NGC 4365 contains few GCs beyond $20$ arcmin and that the sample is complete and free of contamination. We determine the total number of GCs in NGC 4365's GC system to be $6450\pm 110$. This is likely to be a lower limit and doubles the expected number of GCs in NGC 4365 from the value of $3246\pm598$ calculated by \citet{VCS15}. They use the narrow field of view of  one HST/ACS and three HST/WFPC2 pointings to derive an extrapolated GC surface density profile that is not publicly available. We suggest that the surface density profile that \citet{VCS15} fit to the GC candidates underestimated the radial extent of NGC 4365's GC system. 

These total GC numbers convert to specific frequency \citep{Ha81} values of $S_N=3.86\pm0.71$ \citep{VCS15} and $S_N=7.75\pm0.13$ determined from this work for $M_V=-22.31$. Several other large elliptical galaxies with similar $M_V$ have specific frequency values comparable to the $S_{N}$ measured here. NGC 1407 has $S_{N}=7.98\pm0.87$, NGC 1399 has $S_{N}=6.72\pm0.81$ \citep{Spi08} and \citet{VCS15} find $S_{N}=5.2\pm1.4$ for M84. Fitting a power law to the radial density beyond the S\'{e}rsic effective radius we find a slope (i.e.\ power-law exponent) of $-1.21\pm 0.03$ to allow comparisons with analysis of other galaxies in the literature.

\begin{figure}
% \begin{minipage}[t]{0.45\textwidth}\centering
  \includegraphics[width=0.47\textwidth]{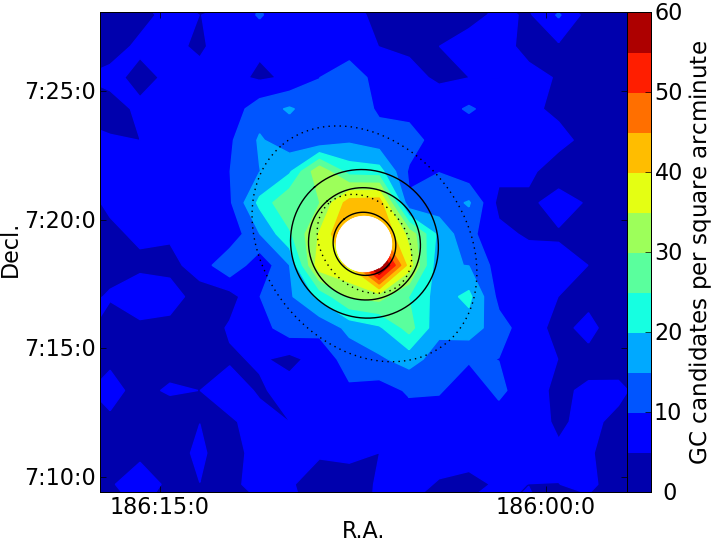}
% \end{minipage}
% \hspace{0.1mm}
% \begin{minipage}[t]{0.5\textwidth}\centering
%  \includegraphics[width=1.0\textwidth]{acs_ellip_dens.png}
% \end{minipage}
   \caption{Surface density of GC candidates in two dimensions from S-Cam data. %S-Cam GC candidates are plotted on the left and ACS GC candidates on the right. 
   The colour ranges from red to blue indicating decreasing GC surface density. Ellipses showing the ellipticity and position angle of the GC system are plotted with dotted black lines and galaxy light isophotes are plotted for comparison in solid black lines. The GC system is much more elongated than the starlight.}
   \label{fig:dens2d}
\end{figure}
\begin{figure}
 \includegraphics[width=0.47\textwidth]{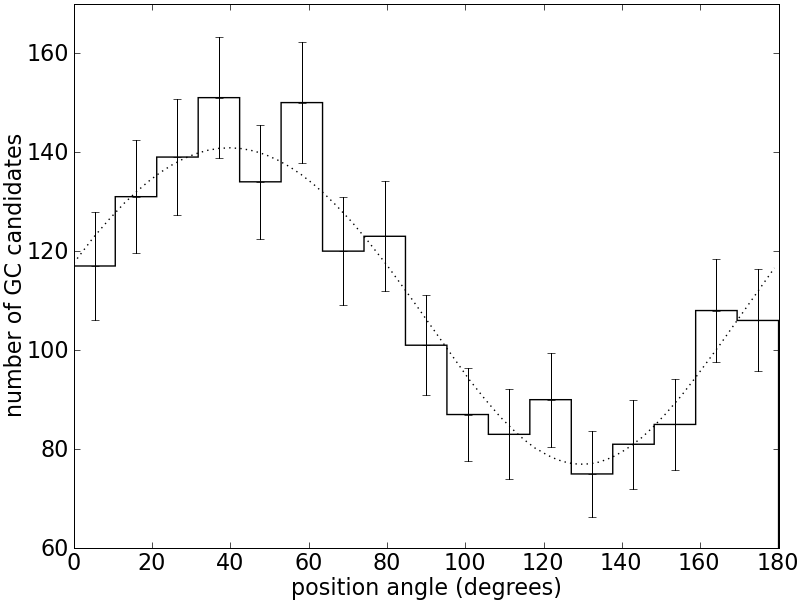}
 \caption{Histogram of the azimuthal distribution of GC candidates. Poisson errors are shown on the histogram and the dotted line with best fit ellipticity and position angle is overplotted. See text for fitting and parameter details.}
 \label{fig:gcpa}
\end{figure}

By binning the GC candidate positions into R.A.\ and Dec.\ defined squares we derive a two-dimensional density distribution for the GC candidates. This is shown in Figure \ref{fig:dens2d} along with selected galaxy surface brightness isophotes from the $i'$ filter as determined by the IRAF task ELLIPSE. \citet{Mc94} use 
\begin{align}\sigma(R,\theta)=kR^{-\alpha}[\cos^{2}(\theta-PA) \mspace{150mu} \notag \\
+(1-e^2)^{-2}\sin^{2}(\theta-PA)]^{-\alpha/2}+bg 
\end{align}
to fit for the position angle ($PA$) and ellipticity ($e$) of the GC system of M87. They fit to the density of GCs as a function of azimuthal distribution and we follow the same procedure. We use the previously determined power law exponent ($\alpha$) and the fitted background value ($bg$), allowing the normalisation constant (k) as well as the position angle and ellipticity to vary.
We find the position angle of the GC system to be $39.7\pm2.2^{\circ}$ and the ellipticity to be $0.63\pm0.02$, see Figure \ref{fig:gcpa}.
%From this it is possible to derive the position angle and ellipticity of the GC system of NGC 4365. \textit{Currently Figure \ref{fig:dens2d} shows ellipses overlaid by hand and at varying ellipticities but constant position angle but I would like it to be done with iraf ellipse eventually}. The position angle of the GC system is consistent with that of the galaxy light (see Section 6) but  
The GC system of NGC 4365 has a similar position angle to the galaxy light ($\sim 42^{\circ}$) but is clearly much more elliptical than the starlight of NGC 4365 (which has an ellipticity of $\sim 0.25$). We see qualitative agreement with this analysis in Figure \ref{fig:dens2d}.

\subsubsection{Colour Distribution}

\begin{figure}
 \includegraphics[width=0.47\textwidth]{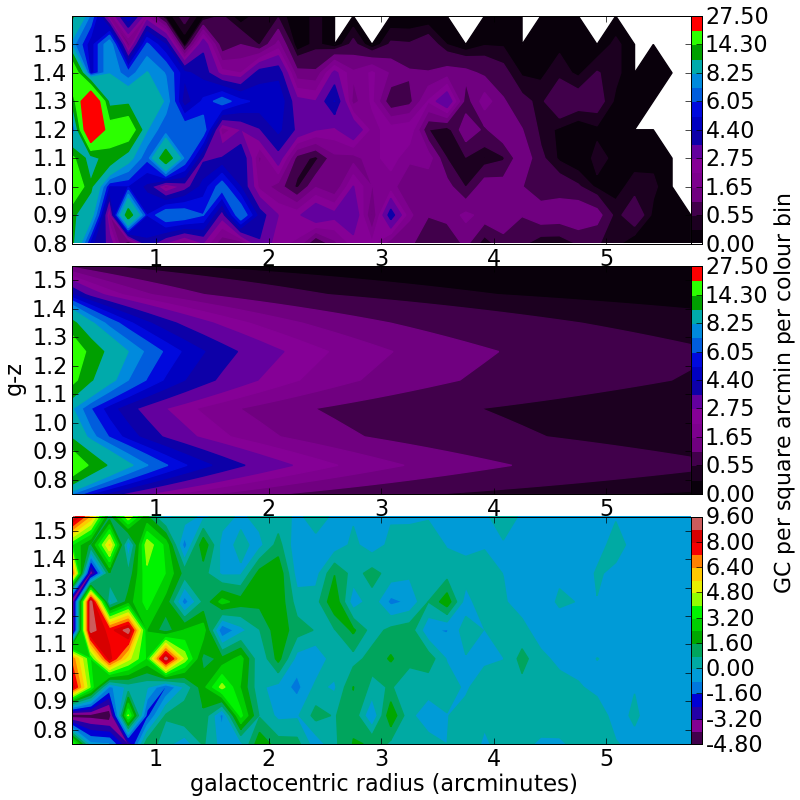}
 \caption{GC density with colour and galactocentric radius for the ACS data. The colour scale indicates object number per square arcmin per colour bin ranging from black at low density to red at high density. The plot is constrained to objects with radii of $0.02$ to $5.8$ arcmin. \textbf{Top}: Measured GC density with colour and radius is plotted. \textbf{Centre}: The model of red and blue GC density calculated using the Gaussian positions and widths from the KMM results for the two subpopulations at large radii ($r>3.5$ arcmin). The proportion of GCs in each Gaussian distribution is taken from the KMM results for the whole sample and the S\'{e}rsic radial density profile used is calculated in Section 6.1.1. This was scaled to have a similar object density to the data at large radii. \textbf{Bottom}: The model in the central panel has been subtracted from the density distribution in the top panel to show an approximate distribution of GCs that are not well modelled by a bimodal colour distribution of GCs. This distribution is not a robust measure but it does show indications of significant numbers of intermediate colour GCs ($g-z \sim 1.1$), particularly at small radii ($\lesssim 2$ arcmin).}
 \label{fig:radcoldensV}
\end{figure}

\begin{figure*}
 \includegraphics[width=1.0\textwidth]{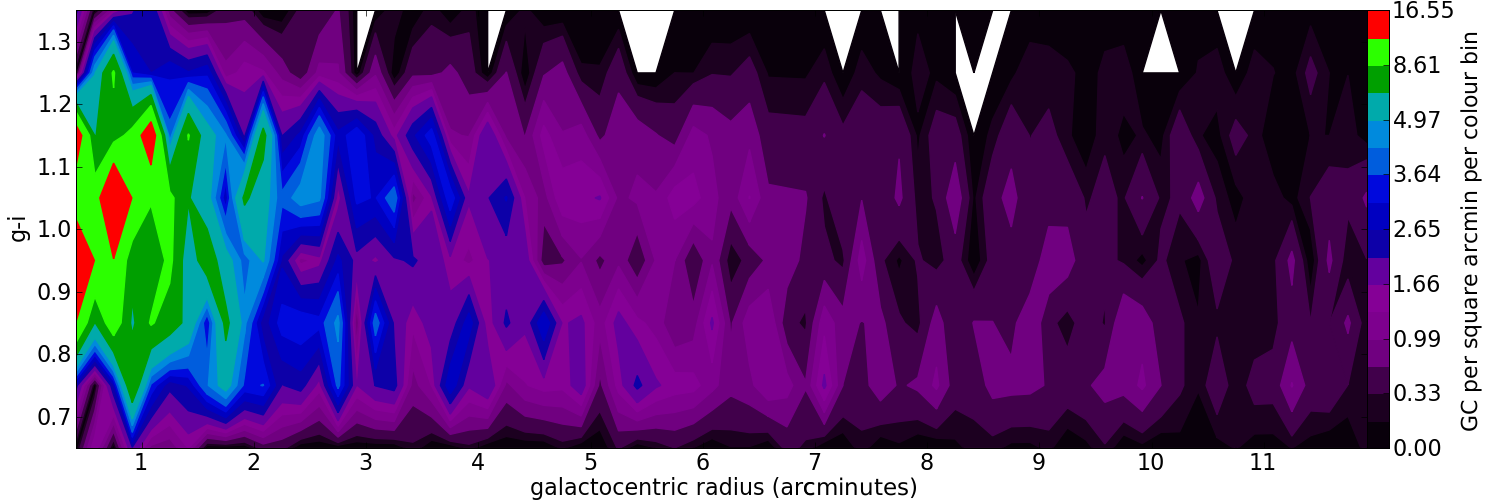}
 \caption{GC density with colour and galactocentric radius for the S-Cam data. The colour scale indicates object number per square arcmin per colour bin ranging from black at low density to red at high density. The plot is constrained to objects with radii of $0.3$ to $12$ arcmin. There is indication of intermediate colour ($g'-i' \sim 1.05$) GCs at small radii, similar to the ACS data, but it is not clear whether this intermediate colour subpopulation is evident at larger radii.}
 \label{fig:radcoldensN}
\end{figure*}

%Without any prior assumptions about the properties of GC subpopulation in NGC 4365 we can only make a qualitative assessment of the surface density of GCs of particular colours. 
Before performing a quantitative analysis of the surface density characteristics of GC subpopulations in NGC 4365 we can make a qualitative assessment of colour-radius substructure by examining the surface density of GCs in colour-radius space. This is shown in Figures \ref{fig:radcoldensV} and \ref{fig:radcoldensN}. At galactocentric radii larger than $2$ arcmin in the top panel of Figure \ref{fig:radcoldensV} the blue and red GC subpopulations of NGC 4365 are visible as distinct sequences at $g-z\sim0.9$ and $\sim1.3$ respectively. At radii smaller than $2$ arcmin and possibly between $2.5$ and $3.5$ arcmin there is evidence of a third subpopulation of GCs intermediate in colour. Note the clear overdensity of GCs candidates at intermediate to red colours and within $1$ arcmin in both Figures \ref{fig:radcoldensV} and \ref{fig:radcoldensN}. These data confirm a similar observation by \citet{La05} where evidence of intermediate colour GCs were found at radii smaller than $1.25$ arcmin. We are able to extend the analysis of the GC colour substructure with radius to $5.5$ arcmin with HST/ACS imaging (see Figure \ref{fig:radcoldensV}) and to $11.5$ arcmin with S-Cam imaging (see Figure \ref{fig:radcoldensN}). Beyond $3$ arcmin in Figure \ref{fig:radcoldensN} the colour distribution seems to be dominated by 2 colour modes but there is some indication of GCs at intermediate colours.

Recently, observers have found gradients in the peak colour of the GC subpopulations with galactocentric radius \citep[see][]{Ha09a,Ha09b,FCS10,Fo11}. Both blue and red subpopulations' peak colour shifts to bluer colours at larger radii. \citet{Fo11} find the most extreme gradient to date in the galaxy NGC 1407. If the red subpopulation of NGC 4365 were similar to that of NGC 1407 then the peak colour for the red subpopulation would change from $g'-i'\sim 1.19$ at the galaxy centre to $g'-i'\sim 1.06$ at $10$ arcmin, and the red subpopulation would have intermediate colours at large radii. In the bottom panel of Figure \ref{fig:radcoldensV} and in Figure \ref{fig:radcoldensN} an overdensity of GC candidates with intermediate colours is visible at very small radii that cannot be accounted for by a gradient in the peak colour of the red subpopulation. We note that if this intermediate colour overdensity at small galactocentric radii is a distinct subpopulation and there is a gradient in the red subpopulation of NGC 4365 then statistical tests for bimodality or trimodality based on colour distribution alone might not be able to distinguish between these two subpopulations conclusively.  
%\textit{I'll include here a paragraph about the possibility of a green subpopulation at very small radii combing with a red subpopulation that becomes greener at larger radii. I had a look at it with Lee and without units for the relation you mentioned and a detailed look at what sort of radii such a relation is applicable I couldn't proceed further.} LEE SUGGESTS IT GO ELSEWHERE ...

\subsection{Radial colour gradients}
\begin{figure}\centering
 \includegraphics[width=0.47\textwidth]{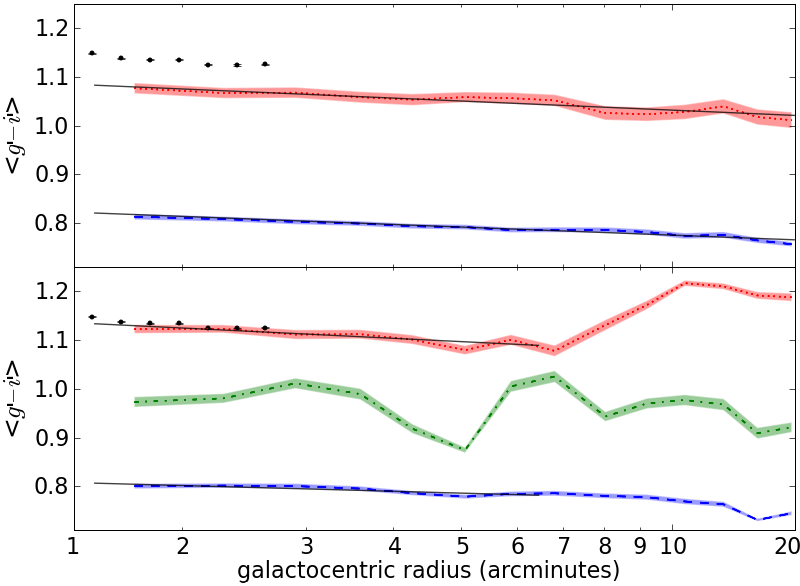}
 \caption{Rolling peak colour values found by the KMM code plotted against galactocentric radius. The bimodal case is plotted in the top panel and the trimodal case in the bottom panel. S-Cam ($g'-i'$) rolling peak values are plotted for blue (dashed), green (dash-dotted) and red (dotted) modes in the multimodal fits. Galaxy $g'-i'$ colours are plotted as black dots with errorbars and the solid black lines show best fit gradients in colour-log radius space.}
 \label{fig:colgrad}
\end{figure}
Several groups have recently found shallow but significant gradients in the mean colour of GC subpopulations \citep[see][]{Ha09a,Ha09b,FCS10,Fo11}. The radial extent of the Subaru/S-Cam imaging makes this an excellent data set to determine whether there are gradients present in the GC subpopulations of NGC 4365. The presence of an intermediate colour subpopulation will influence the gradient determined and therefore we consider the case of a bimodal and a trimodal distribution separately. The heteroscedastic KMM algorithm was run on a rolling sample of GC candidates and the peak value determined for both bimodal and trimodal instances for the S-Cam data set. The rolling sample consisted of 360 GC candidates input to KMM at each step and between each step the 120 candidates with lowest galactocentric radius were replaced with the 120 candidates closest to, but at least as far as, the furthest candidate in the previous sample. A linear relation was fitted to the S-Cam peak colours, to $14.5$ arcmin in the bimodal case and $7$ arcmin in the trimodal case, i.e.\
\begin{equation}
g-i=a+b\,R(\mathrm{arcmin})
\end{equation} 
Results are tabulated in Table \ref{tab:lin}. The linear blue gradient values are consistent within errors but the linear red gradient is significantly steeper in the trimodal case. The green gradient measured in the trimodal case is consistent with zero.
\begin{table}\centering
 \begin{tabular}{ccc}
 \hline
\textbf{Subpopulation} & \textbf{Bimodal Case}  & \textbf{Trimodal Case} \\
& (dex per arcmin) & (dex per arcmin) \\ \hline \hline
 Blue & $-0.0037\pm 0.0002$ & $-0.004\pm 0.001$ \\
 Green & - & $0.001\pm0.012$ \\
 Red & $-0.0047\pm 0.0005$ & $-0.009\pm 0.002$ \\ \hline
 \end{tabular}
 \caption{GC colour gradients for the linear fit to colour with radius in arcmin (Equation 5) for S-Cam data.} %($\frac{\Delta(g-i)}{\Delta(R)}$)
 \label{tab:lin}
\end{table}

\begin{table}\centering
 \begin{tabular}{ccc}
 \hline
\textbf{Subpopulation} & \textbf{Bimodal Case}  & \textbf{Trimodal Case} \\
& (dex per dex) & (dex per dex) \\ \hline \hline
 Blue & $-0.055\pm 0.004$ & $-0.04\pm 0.01$ \\
 Green & - & $-0.01\pm0.10$ \\
 Red & $-0.062\pm 0.008$ & $-0.07\pm 0.02$ \\ \hline
 \end{tabular}
 \caption{GC colour gradients for the logarithmic fit to colour with normalised radius (Equation 6) for S-Cam data.} % ($\frac{\Delta(g-i)}{\Delta(\mathrm{log}(R/R_e))}$)
 \label{tab:log}
\end{table}
We also fit a line to the logarithm of the radius normalised by $R_e$ using $R_e=2.1$ arcmin, i.e.\
\begin{equation}
g-i=c+d\, \mathrm{log}(R/R_e)
\end{equation}
Results are tabulated in Table \ref{tab:log}. Here both blue and red gradients are consistent within errors for the bimodal and trimodal cases. The green subpopulation does not have a measurable gradient in this case either. \citet{FCS10} measure colour gradients for NGC 4365 GCs using the central ACS pointing in the bimodal case, finding a blue slope of $-0.056\pm0.024$ and a red slope of $-0.033\pm0.021$ dex per dex. The radial range of their measurements do not overlap with ours and we see that the radial colour gradient of the blue subpopulation remains constant when moving to larger galactocentric radii while the radial colour gradient of the red subpopulation steepens significantly with increasing galactocentric radius. We use the \citet*{Le10b} empirical relationship between GC colour and metallicity to give a metallicity gradient, i.e.\ %($\frac{\Delta[\mathrm{Fe/H}]}{\Delta(\mathrm{log}(R/R_e))}$) 
\begin{equation}
\Delta[\mathrm{Fe/H}]=3.48\Delta(g-i)
\end{equation}
\begin{table}\centering
 \begin{tabular}{ccc}
 \hline
\textbf{Subpopulation} & \textbf{Bimodal Case}  & \textbf{Trimodal Case} \\
& (dex per dex) & (dex per dex) \\ \hline \hline
 Blue & $-0.19\pm 0.01$ & $-0.13\pm 0.03$ \\
 Red & $-0.22\pm 0.03$ & $-0.26\pm 0.06$ \\ \hline
 \end{tabular}
 \caption{Metallicity gradients for the logarithmic fit to colour with normalised radius (Equations 6 \& 7) for S-Cam data.} % ($\frac{\Delta[\mathrm{Fe/H}]}{\Delta(\mathrm{log}(R/R_e))}$)
 \label{tab:mtl}
\end{table}
The gradient of the mean metallicity with galactocentric radius is shallower in the trimodal case for the blue subpopulation but for the red subpopulation the gradients in the bimodal and trimodal cases are consistent. We note that the rolling peak colour found by KMM is only consistent with the galaxy colour in the trimodal case (see Figure \ref{fig:colgrad}). 

We conclude that the NGC 4365 GC subpopulations show a shallow but significant colour gradient with galactocentric radius, regardless of whether the GC system is interpreted as being bimodal or trimodal. \citet{Fo11} tabulate metallicity-radius gradients in units of dex per dex for several galaxies on which this measurement has been done. The gradients measured for NGC 4365 are steeper than the mean, but not the steepest measured. Results vary from $-0.10$ to $-0.38$ for the blue subpopulation and from $-0.10$ to $-0.43$ for the red subpopulation in \citet{Fo11} Table 1. These gradients explain why statistical analyses of the colour distribution (collapsed in radius) of NGC 4365's GC system give inconclusive results.% and probably exclude the green ellipticity estimate. % NOT REALLY TRUE A trimodal interpretation of NGC 4365's GC system is supported by the fact that the mean red GC colours are only consistent with the galaxy light if trimodality is assumed, \citet{Br06} note that red GC follow that galaxy light colour profile for almost all giant elliptical galaxies.

\subsection{Magnitude colour trends}
\begin{figure*}\centering
 \includegraphics[width=0.80\textwidth]{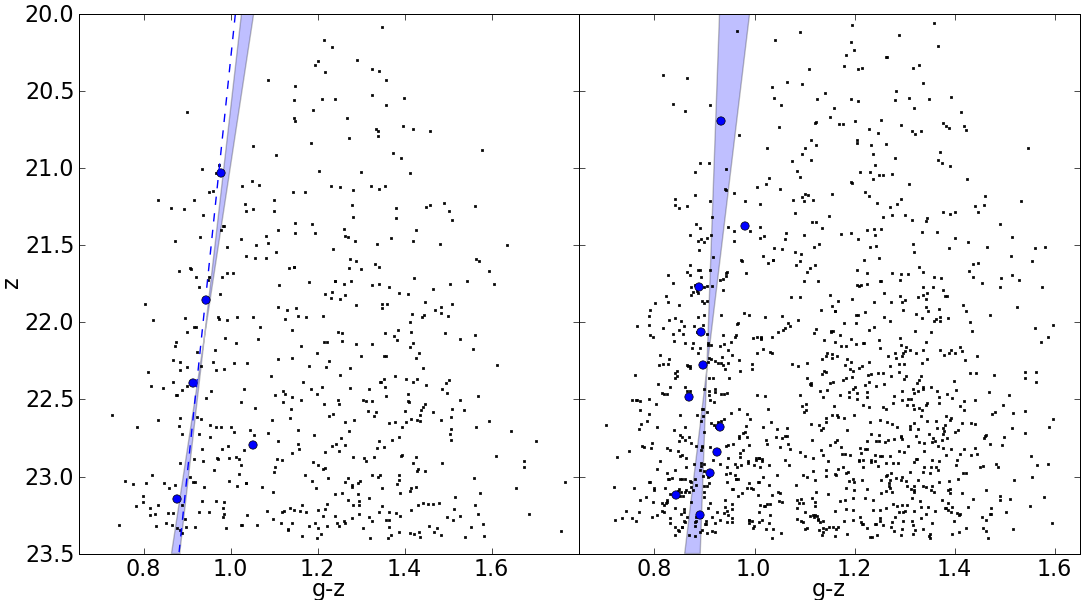}
 \caption{The colour-magnitude relation for blue GC candidates found in the HST/ACS photometry. The left panel shows GC candidates found in the central ACS pointing and the right panel shows candidates found in all eight ACS pointings. The KMM mean value for the blue GCs is overplotted with blue points. The one sigma range for the linear fit to those values is overplotted with a blue shaded region on both panels. %Also on both panels a dotted line with zero gradient is plotted for comparison. 
 On the left panel the gradient the slope found by \citet{VCS14} for galaxies between $-21.7<M_B<-21$ is overplotted with a dashed line.}
 \label{fig:bluetilt}
\end{figure*}

In the determination of the magnitude colour trends we only investigate the blue subpopulation of the GC system, we search for a `blue tilt' \citep{St06}. We do this in order to decide whether the blue tilt could be the cause of the intermediate colour overdensity seen in Figures \ref{fig:cmda}, \ref{fig:radcoldensV} and \ref{fig:radcoldensN}. The bimodal KMM mean colour values at various magnitudes were determined by dividing the GC sample into magnitude subsets with equal numbers. The mean $z$ magnitude of GC candidates in each subset is plotted against the blue heteroscedastic Gaussian mean value found by KMM in Figure \ref{fig:bluetilt}. We did this analysis for the central HST/ACS pointing as well as for the full set of eight HST/ACS pointings and found that both cases show a significant gradient but that the slope is steeper in the central region ($-0.0477\pm0.0018$) than for all pointings combined ($-0.024\pm0.013$). \citet{VCS14} use the central ACS pointing and combine GCs for all galaxies in the range  $-21.7<M_B<-21$ to determine a slope of $-0.037\pm0.004$ using the same KMM method. In Figure \ref{fig:bluetilt} we see that this lies very close to our fitted line. They also find a steepening slope when GCs closer the the galaxy centre are used.

In Figure \ref{fig:cmda}, indication of a clump of intermediate colour/green objects is seen around $g-z\sim 1.1$ and $22>z>20$ and in the bottom panel of Figure \ref{fig:radcoldensV} a clear overdensity is seen inside 2 arcmin and around $g-z\sim 1.1$. The colour magnitude slope for the blue GC candidates is not steep enough to cause either of these overdensities.

\subsection{Quantifying the GC subpopulations}
We compare NGC 4365's GC colour distribution with that of other Virgo cluster galaxies, using the ACS Virgo Cluster Survey (VCS) photometry \citep{VCS9}. They fit a homoscedastic bimodal distribution using Kaye's Mixture Model (KMM) algorithm and obtained mean $g-z$ colour values of 0.98 and 1.36 for the blue and red subpopulations of NGC 4365 respectively, as well as a width of 0.15 for both subpopulations. %These values for the mean colours of the subpopulations fall close to the g-z vs.\ $M_{B}$ relations as determined in \citet{VCS9} when the distance of NGC 4365 is assumed to be similar to that of the centre of the Virgo cluster (16.5 Mpc). 
They also fit linear relations between mean $g-z$ colour and absolute galaxy magnitude, $M_{B}$, for both GC subpopulations, assuming all Virgo Cluster galaxies have the same distance. The relations predict mean $g-z$ colour values of $0.98\pm0.06$ and $1.40\pm0.08$ for the blue and red subpopulations when the conventional distance for NGC 4365 (23.1 Mpc) is used. They concluded that aside from having a large number of red GCs, NGC 4365 has an unremarkable colour distribution compared to their sample \citep{VCS9}. %We note there is some inherent inaccuracy in the relations calculated by \citet{VCS9} because they used a mean Virgo Cluster distance rather than exact distances to the galaxies, which is 6.6 Mpc too close for NGC 4365 \citep{VCS13}. %but we assume that this one galaxy has little impact on the derived relations.

Fitting a bimodal homoscedastic distribution to NGC 4365's GCs, with KMM, using all eight available ACS pointings we find similar values to the ones found by ACS VCS (i.e.\ $g-z=$ 0.94 and 1.30 for the blue and red peak colours and a width of 0.12). However, on visual inspection the fit to the GC distribution is skewed to redder colours when compared to the actual colour distribution. We also fit both bimodal and trimodal heteroscedastic distributions to the GC colour distributions, obtaining mean blue and red values of $g-z=$ 0.89 and 1.25 in the bimodal case, and 0.88, 1.12 and 1.34 in the trimodal case. In neither case does the mean blue colour lie close to the $g-z$ vs.\ $M_{B}$ relation (values are 1.4$\sigma$ and 1.6$\sigma$ away from the relation) but only in the trimodal case does the red mean colour lie close to the \citet{VCS9} $g-z$ v.s.\ $M_{B}$ relation, where the value is 0.7$\sigma$ away from the relation instead of 1.8$\sigma$ away. This is a motivation to describe the GC system of NGC 4365 in terms of three rather than two subpopulations.
%This is a $1\sigma$ indication that a trimodal distribution is a more accurate description of the overall colour distribution of NGC 4365's GCs. Because we have better agreement with the g-z vs.\ $M_{B}$ relations in the case where we fit a trimodal colour distribution we suggest that a trimodal distribution, containing an intermediate colour third subpopulation, is possibly a more accurate description of the colour distribution of NGC 4365's GCs.

%\citet{VCS9} also determine a relationship between subpopulation distribution width and $M_{B}$ by determining the width of the subpopulations in a stacked sample of GCs from all galaxies in a $M_{B}$ range. We note that the ACS Virgo Cluster Survey widths are probably overestimates for single galaxies as the distributions measured in their sample are superpositions of several galaxies with small but significant scatter in their mean subpopulation colour values and are therefore not comparable with our result for NGC 4365.
\begin{figure}\centering
 \includegraphics[width=0.47\textwidth]{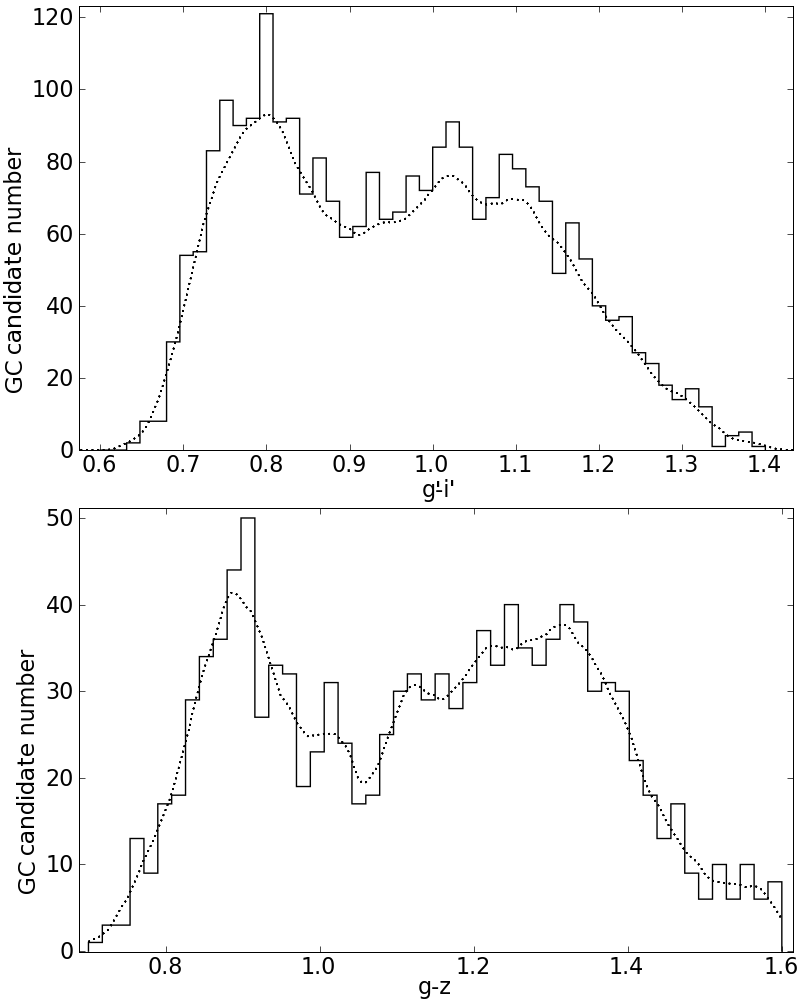}
 \caption{Colour distribution of GC candidates at all radii for S-Cam (top) and ACS (bottom) brighter than the turnover magnitude ($z = 23.4$ and $i' = 23.6$). The Epanechnikov kernel smoothing of the colour distribution is plotted with a dotted line, the smoothing width used is the average colour error (0.04 for both ACS and S-Cam). At least three peaks can be seen.}
% and the outer two peak positions were used to define the blue and red GC colour distributions, see Section 3.3}
 \label{fig:hist}
\end{figure}

%\subsubsection{Statistical tests and comparisons with other giant elliptical galaxies}
From this point forward we assume that the GC system of NGC 4365 contains three subpopulations, a blue, green (intermediate colour) and red subpopulation. There are radial gradients in colour and a blue tilt but in neither case are they enough to explain the strong overdensity seen at intermediate colours in both Figures \ref{fig:radcoldensV} and \ref{fig:radcoldensN}. In addition, there is better agreement with the colour - galaxy luminosity relations of \citet{VCS9} to the other Virgo cluster galaxies and a near perfect match between galaxy light and mean red subpopulation colours if we assume that NGC 4365 has three subpopulations.
%the peak and width of each of the three Gaussian distributions making up the trimodal distribution as well as

%We attempted to determine the significance with which a trimodal distribution is preferred over a more common bimodal distribution. We used several statistical tests including the likelihood ratio test statistic from the KMM algorithm, the  Kolmogorov-Smirnov (KS) test and the minimum Chi-Squared ($\chi^{2}$) test, but results of these analyses are inconclusive. Using the likelihood ratio test statistic from KMM trimodality is preferred over bimodality for NGC 4365's GC but also for NGC 1407s GCs where trimodality is not suspected (data from Spitler et al.\ 2011 in prep. also used in \citealt{Ro09,Fos10,Fo11}). The KS test shows that the actual colour distribution is significantly different from both bimodal and trimodal simulated distributions and the reduced $\chi^{2}$ statistic decreases significantly when a third mode is included in the colour distribution but that mode is very close to the original blue mode, not roughly midway between the blue and red subpopulations (see Appendix A for further details on all three tests). To better investigate the significance of a three subpopulation split relative to a two subpopulation split we divide the GC distribution into three separate subpopulations with the aim of determining if the properties of the intermediate colour (or green) GC candidates are significantly different from the properties of either the red or blue candidates. 

In the trimodal case, the green GCs would likely `contaminate' the blue and red subpopulations to all except the bluest and reddest colours. A simple split based on colour would leave a high percentage of green GCs in both the blue and red subpopulations and visa versa. Therefore, we formulate a probability that a GC belongs to either a blue, green or red subpopulation based on its colour and galactocentric radius. Note that red (metal rich) GCs are generally found to be more centrally concentrated than blue (metal poor) GCs \citep{Br06} thus GC galactocentric radius is an important factor in the probability calculation. We assume that each of the three subpopulations are fairly well described by a Gaussian in colour (without skewness or kurtosis) and that the peak and width of the Gaussian is independent of galactocentric radius (i.e.\ no radial colour gradient) but make no further assumptions about the radial distribution of any of the three subpopulations. Given evidence of radial colour gradients we note that this assumption is strictly inaccurate. We make this assumption because colour gradients are shallow compared to the width of the GC colour distributions and the calculation of gradient values contain other significant assumptions about the radial distribution of the subpopulations that we wish to avoid. The following paragraphs describe how we assign a probability that a GC is blue, green or red (henceforth blue-green-red probability). %dependent on the colour and galactocentric radius of the GC. %Henceforth we refer to the possible intermediate colour subpopulation as the green subpopulation.

\begin{table}\centering
 \caption{Gaussian values for the blue and red GC distributions as determined from an Epanechnikov smoothing kernel of the colour distribution.}
 \begin{tabular}{c|c|c|c|c|} \hline
 & \multicolumn{2}{|c|}{\textbf{ACS (g-z)}}  & \multicolumn{2}{|c|}{\textbf{S-Cam (g'-i')}} \\ 
 & peak & width & peak & width \\ \hline \hline
Blue & 0.89 & 0.07 & 0.80 & 0.07\\
Red & 1.32 & 0.12 & 1.13 & 0.10 \\ \hline
 \end{tabular}
 \label{tab:gvals}
\end{table}
 
\begin{figure*}
 \includegraphics[width=1.0\textwidth]{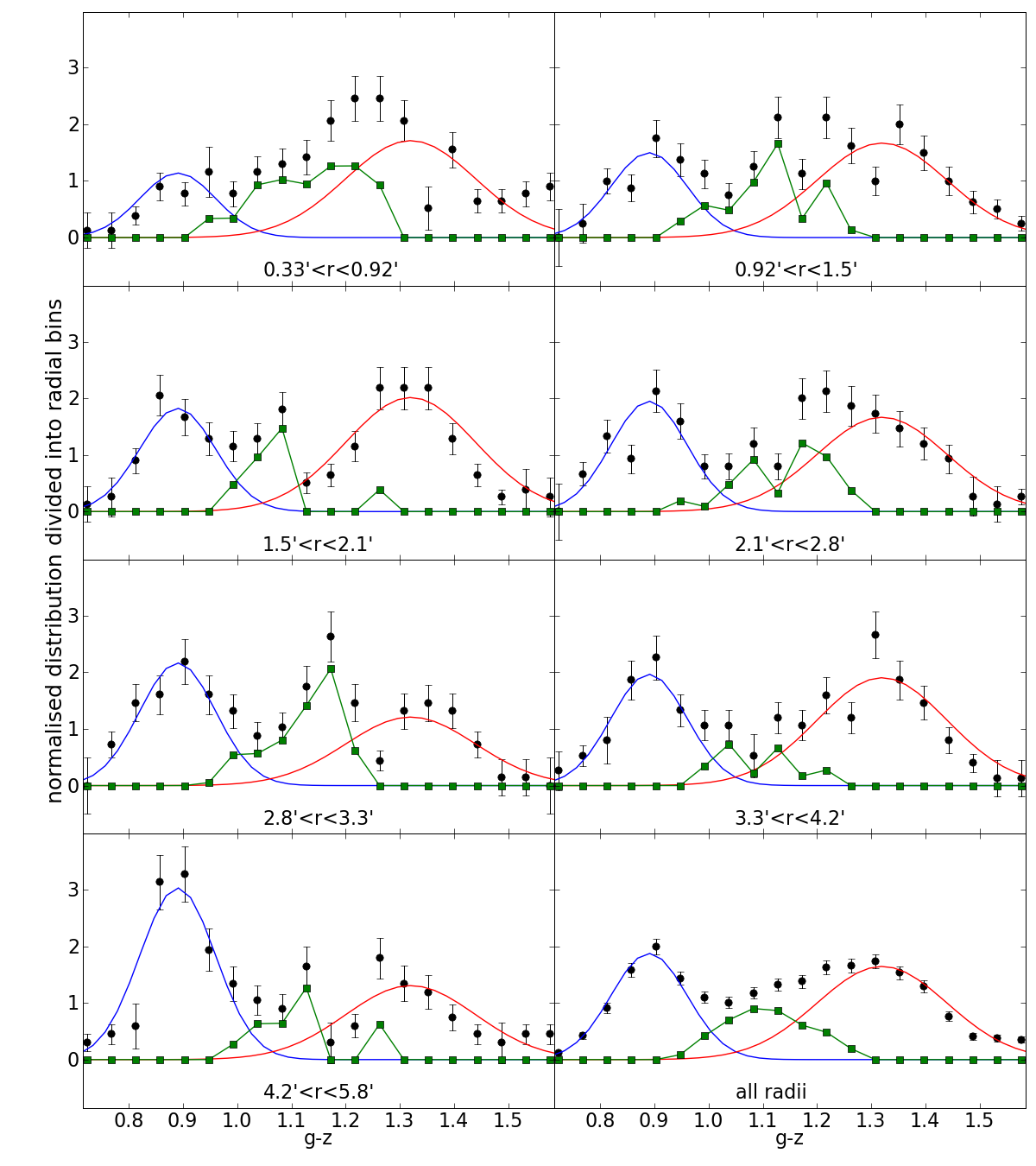}
 \caption{The normalised distribution of ACS GCs with colour in the seven radial bins (shown in arcminutes on each plot) and at all radii (lower right panel). Histogram heights are shown as black circles with errorbars representing Poissonian errors. Fitted Gaussian distributions are plotted in blue and red for the respective subpopulations and the excess GCs (calculated only in the region between the blue and red peaks) are plotted as green squares. The peak and width parameters for the Gaussian distributions fitted in each case are constant: $\mu_{b}=0.89$, $\sigma_{b}=0.07$, $\mu_{r}=1.32$ and $\sigma_{r}=0.12$, with the normalisation allowed to vary.}
 \label{fig:acsradbin}
\end{figure*}

\begin{figure*}
 \includegraphics[width=1.0\textwidth]{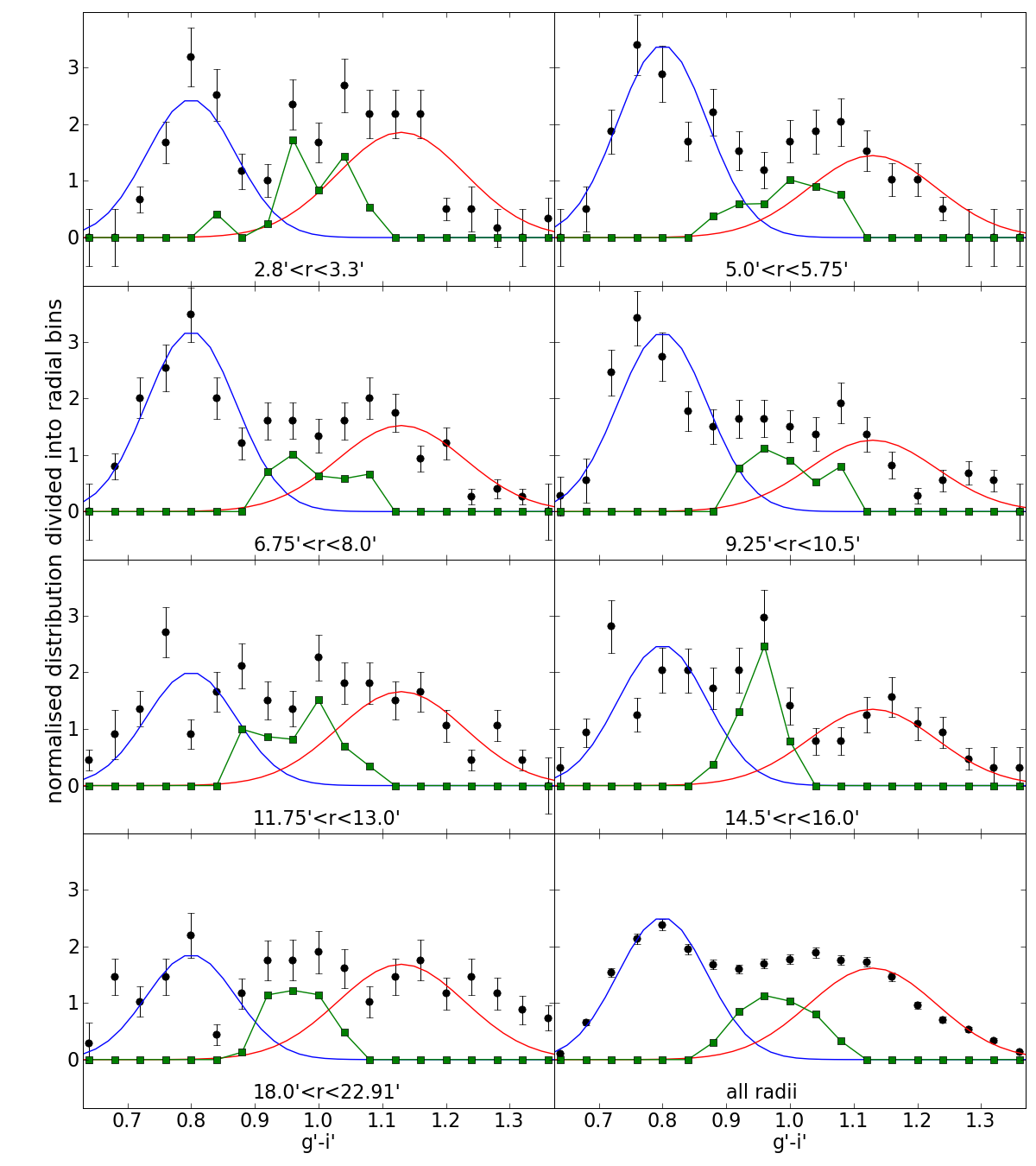}
 \caption{The normalised distribution of S-Cam GCs with colour in seven radial bins (shown in arcminutes on each subplot) and at all radii (lower right panel). Histogram heights are shown as black circles with errorbars representing Poissonian errors. Fitted Gaussian distributions are plotted in blue and red for the respective subpopulations and the excess GCs (calculated only in the region between the blue and red peaks) are plotted as green squares. The peak and width parameters for the Gaussian distributions fitted in each case are constant: $\mu_{b}=0.80$, $\sigma_{b}=0.07$, $\mu_{r}=1.13$ and $\sigma_{r}=0.10$, with the normalisation allowed to vary.}
 \label{fig:scamradbin}
\end{figure*}

The peak of the blue Gaussian distribution was determined from the bluest peak of an Epanechnikov kernel smoothing of the colour distribution \citep{Sil86} and the width of the blue Gaussian distribution was determined from the colour at which 68 per cent of the GCs bluewards of the peak were included in the distribution (assuming that green GCs would not be present in significant numbers at such blue colours). See Figure \ref{fig:hist} for the S-Cam colour distribution and kernel smoothing. The peak position and width of the red distribution were calculated similarly and all values are shown in Table \ref{tab:gvals}. As seen in the bottom right panels of Figures \ref{fig:acsradbin} and \ref{fig:scamradbin}, where GCs at all radii are included for each sample, the Gaussian distributions for blue and red subpopulations determined in this way are a reasonable fit to the extreme blue and red ends of the GC colour distribution.

%The peak of both the blue and red distributions were determined by analysis of a histogram with very narrow bins ($\le 10^{-4}$ in colour). There is a lot of scatter in a histogram with such narrow bins but it ensures that the binning of the data does not alter the position of the peak calculated. The blue peak in the GC colour distribution is seen as a group of spikes at the extreme blue end of the colour distribution and we average the positions of all the spikes in the group to determine the position of the blue peak. The width of the blue Gaussian distribution was determined from the colour at which 68\% of the GCs bluewards of the peak were included in the distribution (assuming that green GCs would not be present at such blue colours). The Gaussian values found for the S-Cam catalogue were $\mu_{blue}^{S-Cam}=0.80$ and $\sigma_{blue}^{S-Cam}=0.07$ and those found for the ACS catalogue were $\mu_{blue}^{ACS}=0.89$ and $\sigma_{blue}^{ACS}=0.07$. The peak position and width of the red distribution were calculated similarly and found to be $\mu_{red}^{S-Cam}=1.13$, $\sigma_{red}^{S-Cam}=0.10$ and $\mu_{red}^{ACS}=1.32$, $\sigma_{red}^{ACS}=0.12$ for S-Cam and ACS respectively. 
To assess the number of blue, green and red GCs as a function of colour and radius, the GCs in each sample (ACS or S-Cam) were split into radial bins of roughly equal object number and for each radial bin blue and red Gaussian distributions were fitted to the normalised colour histogram. Peak and width values were held constant across radial bins, as determined from the whole sample (see Table \ref{tab:gvals}), and the normalisations of the distributions were fitted separately for each radial bin. This was a $\chi^2$ minimization fit. The blue and red numbers were calculated by adding the GCs in the extreme colour regions (where no significant green `contamination' is expected) to the percentage of the total objects expected to belong to either the blue or red subpopulations (obtained by integrating the blue and red Gaussian distributions over intermediate colours). The green GC numbers were defined by subtracting both the blue and red subpopulation numbers from the total number in the distribution. These values were used to determine a radial surface density profile for each subpopulation (discussed further in Section 6.5.1), which is used to calculate the probability that a GC at any radius or colour belongs to the blue, green or red subpopulations.

The process used to determine the blue-green-red probability of a GC starts by determining the relative number of GCs in each subpopulation at the galactocentric radius of the GC. This is done by comparing the values of the radial surface density profile for each subpopulation. The relative numbers of GCs in each population are used to scale the normalisations of the three Gaussian distributions and comparing the relative values of all three Gaussian distributions at the colour of the GC we calculate the probability of an object being blue, green, or red.
\subsection{Characterising the GC system subpopulations}
\subsubsection{Surface density}

\begin{figure}
 \includegraphics[width=0.47\textwidth]{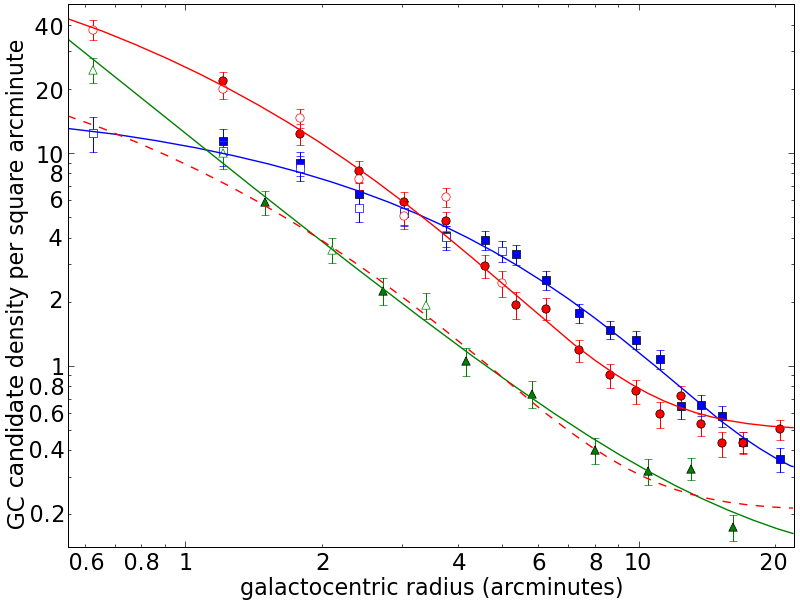}
 \caption{GC radial surface density for blue (squares), green (triangles) and red (circles) subpopulations incorporating ACS (unfilled points) and S-Cam (filled points) GCs. The surface density for both the blue and red GC subpopulations are well fit by a S\'{e}rsic profile added to a background term (shown as blue and red solid lines). The surface density for the green subpopulation is best fit by a power law plus a background value (shown as a green solid line). The red dashed line is a scaled version of the  S\'{e}rsic profile fit to the red subpopulation that is a reasonable fit to all except the inner two green surface density points. See text for further details.}
 \label{fig:spopdens}
\end{figure}
As described in Section 6.4 we counted the blue, green and red GCs in each radial bin assuming that the red and blue GC colour distributions are well fit by Gaussians. In Figure \ref{fig:spopdens} the radial surface density data for all three subpopulations are shown with the best fit surface density model in each case. The red and blue subpopulation distributions are both well fit by a S\'{e}rsic profile added to a background term. The values we found are recorded in Table \ref{tab:colSers}. The background terms were estimated from the outermost radial surface density value. The effective radius, or half number radius, of the blue subpopulation ($7.30$ arcmin) is more than twice as large as the effective radius of the red subpopulation ($3.17$ arcmin) and the S\'{e}rsic n values are consistent within combined errors. The green subpopulation can be fit by the red subpopulation S\'{e}rsic profile if the two innermost density points are omitted (points interior to $1.5$ arcmin) and we fit for $P_{e}$ and the $bg$ (see Table \ref{tab:colSers} for the fitted values). 

\begin{table}\centering
 \begin{tabular}{l|@{}l|@{}l|@{}l|@{}l}
 \hline
 & {\boldmath$P_{e}$}  & $\mathbf{n}$  & $\mathbf{R_{e}}$ & $\mathbf{bg}$ \\
 & ($\mathrm{arcmin}^{-2}$) & & (arcmin) & ($\mathrm{arcmin}^{-2}$) \\ \hline \hline
 Blue & $1.67\pm0.29$ & $1.36\pm 0.19$ & $7.30\pm 0.68$ & $0.25$\\
 Green & $1.71\pm0.17$ & $2.02$ & $3.17$ & $0.21\pm 0.03$ \\
 Red & $4.91\pm0.59$ & $2.02\pm 0.25$ & $3.17\pm 0.19$ & $0.50$\\ \hline
 \end{tabular}
 \caption{S\'{e}rsic profiles fits to the radial surface density of the blue, green and red GC subpopulations. The green subpopulation $n$ and $R_{e}$ values have been fixed to the red subpopulation values.}
 \label{tab:colSers}
\end{table}

We also fit a two parameter power law plus a background term 
\begin{equation}
\rho(R)=\rho_0\mathrm{R}^{\alpha} + bg
\end{equation}
to the entire radial range of the green subpopulation, finding $\alpha=-1.71\pm0.13$, $\rho_0=12.3\pm1.3\mathrm{\hspace{3pt}arcmin}^{-2}$ and $bg=0.10\pm0.05\mathrm{\hspace{3pt}arcmin}^{-2}$.  %The addition of the background values for each of the three subpopulations, ($bg^{blue} + bg^{green} + bg^{red})=(0.25 + 0.10 + 0.48) = 0.73\pm 0.09\mathrm{\hspace{3pt}arcmin}^{-2}$, is consistent within errors with the background value measured for the whole system, i.e.\ $0.89\pm 0.13\mathrm{\hspace{3pt}arcmin}^{-2}$ and the agreement improves when the green background value determined by the S\'{e}rsic fit is used instead of the value determined by the power law. 
Comparing the slope of a power law fit to the blue ($-1.13\pm0.06$) and red ($-1.60\pm0.04$) distributions over a radial range from 0.5 to 11 arcmin (4 to 70 kpc) we can quantitatively verify that the green and red GC subpopulations are more centrally concentrated than the blue GC subpopulation and that the slope of the green and red subpopulations are consistent with each other.

%When a power law is fit to blue $16.5\pm1.6$ and $-1.13\pm0.06$ and red $31.0\pm1.8$ and $-1.60\pm0.04$
\begin{figure}\centering
 \includegraphics[width=0.47\textwidth]{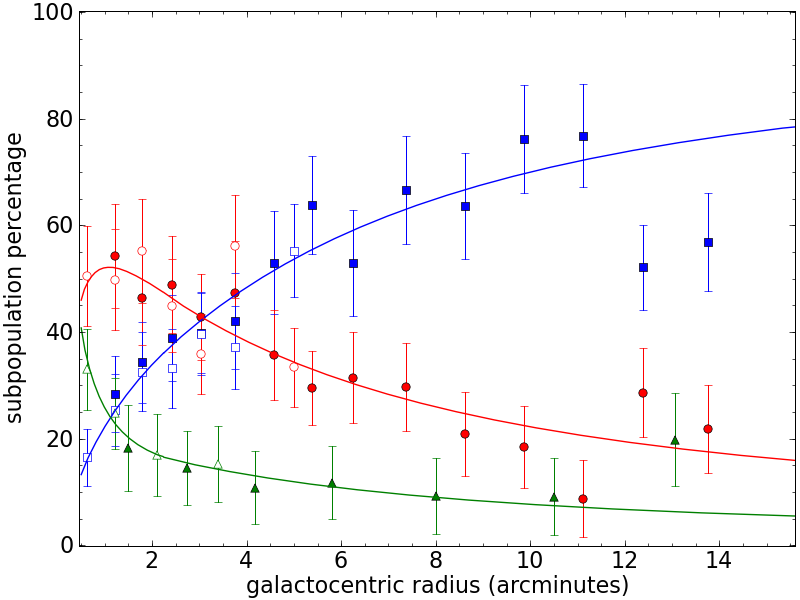}
 \caption{The percentage contribution of each subpopulation to the total number of GC candidates plotted against galactocentric radius. Curves were calculated using the fitted surface density profiles in Figure \ref{fig:spopdens}. Both the data points and plotted profiles have had the respective background values subtracted. Beyond $12$ arcmin the total population numbers are low and the measurement of the percentage contribution of any one subpopulation to the total number becomes unstable. We see that the red subpopulation dominates at small galactocentric radii and the blue dominates at large galactocentric radii while the green subpopulation is only significant at very small galactocentric radii.}
 \label{fig:percent}
\end{figure}
In Figure \ref{fig:percent} the percentage of the total GC candidate number that is attributed to each subpopulation is plotted as a function of galactocentric radius. The data points have been background subtracted and the curves derived from the profile fits in Figure \ref{fig:spopdens} are plotted without the addition of background terms. In the case of the green subpopulation the curve shown is a combination of the power law fit in the inner parts ($R<2.25$ arcmin) and the S\'{e}rsic profile beyond that. We do this to adequately display the behaviour of the intermediate colour surface density both in the inner (power law behaviour) and outer (S\'{e}rsic behaviour) parts, noting that we plot lines to guide the eye and not for analysis. In agreement with previous work \citep[see review by][]{Br06} the red subpopulation dominates the GC system in the inner parts and the blue subpopulation dominates in the outer parts. We also see that the green subpopulation only contributes significantly to the GC system at very small galactocentric radii ($r\lesssim2$ arcmin). This is in agreement with the qualitative analysis of the colour-galactocentric radius properties of the GC system earlier in this work (see Section 6.1.2) and with \citet{La05}.

%\subsubsection{Colour-magnitude variation}
\subsubsection{GC half light sizes}
\begin{figure}\centering
 \includegraphics[width=0.47\textwidth]{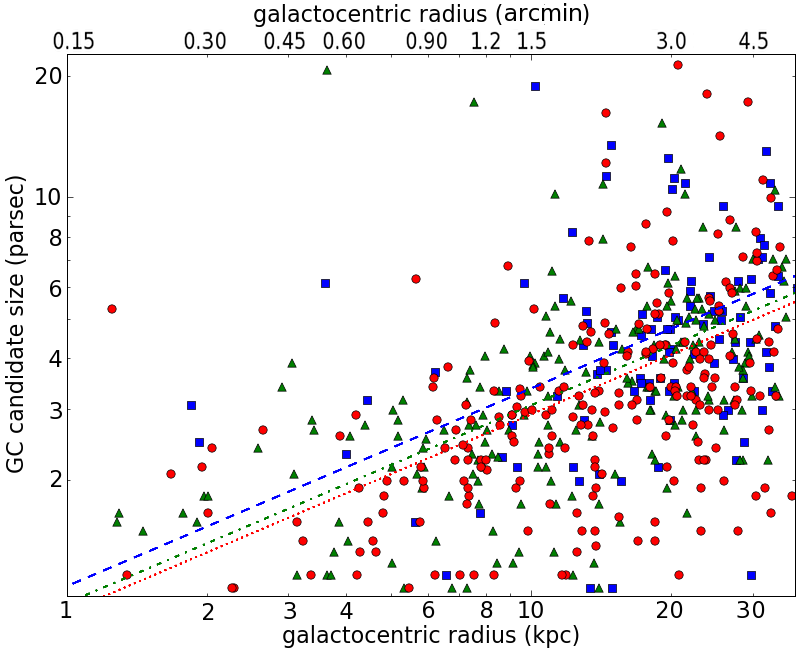}
 \caption{The half light radius ($r_h$) of HST/ACS GC candidates against galactocentric radius for objects brighter than $z=23.4$. The high probability blue, green and red GC subpopulation candidates are plotted with blue squares, green triangles and red circles respectively. There is a trend of increasing GC candidate size with increasing galactocentric radius for all GCs regardless of subpopulation colour. A power law fitted to the whole sample of GC candidates brighter than $z=23.4$ has a slope of $0.49\pm 0.04$ dex per dex. Overplotted with blue dashed, green dash-dotted and red dotted lines are the power law fits for size normalisation (with slope set by whole GC sample) of each high probability subpopulation sample.}
 \label{fig:sizerad}
\end{figure}
 
Here we measure subpopulation sizes (median half light radii) as well as trends in candidate size with galactocentric radius from the ACS photometry. As described in Section 6.4 each GC candidate is assigned a probability of belonging to each subpopulation. For example a candidate $1.28$ arcmin from the galaxy centre with $g-i=0.98$ has probabilities: $\mathrm{p}_{\mathrm{blue}}=18$ per cent, $\mathrm{p}_{\mathrm{green}}=79$ per cent and $\mathrm{p}_{\mathrm{red}}=3$ per cent. Defining the blue and red GCs with a colour probability larger than 95 per cent and green GCs with a probability larger than 80 per cent there are 315, 426 and 491 GC candidates in each of the blue, green and red high probability subsamples. There are no green GC candidates with a probability larger than 95 per cent so we compare medium probability green candidates (80-95 per cent) with high probability blue and red candidates. The aforementioned example is not defined as part of either blue, green or red high probability subsamples.  

We determine the median half light radius of the GCs; 
\begin{align}
r_{\mathrm{h,blue}} = 0.037 \pm^{0.002}_{0.002}\mathrm{\hspace{3pt}arcsec}  \notag \\
r_{\mathrm{h,green}} = 0.027 \pm^{0.001}_{0.001}\mathrm{\hspace{3pt}arcsec} \notag \\
r_{\mathrm{h,red}} = 0.025 \pm^{0.001}_{0.001}\mathrm{\hspace{3pt}arcsec} \notag
\end{align}
These sizes were measured on GCs brighter than $z=23.4$ (109, 223 and 269 blue, green and red high probability subsample GCs) using ISHAPE \citep{Ishape} where the method is described in \citet{St06}. The median sizes correspond to $4.1\pm^{0.3}_{0.2}$, $3.0\pm^{0.2}_{0.1}$ and $2.8\pm^{0.1}_{0.1}$ parsec respectively. Blue GCs have a significantly larger median size than either green or red GCs. However, the sizes of the green GCs are only slightly larger than the red GCs. This is possibly due to distinct characteristic sizes for each subpopulation but could also be explained by a continuous trend in GC size with GC colour \citep[e.g.][]{Jo04}. % within combined $1\sigma$ errors and consistent within combined $2\sigma$ errors. It is not clear whether the sizes of the green GC candidates are different from both red and blue candidates or whether the median sizes of the green subpopulation are slightly larger than the red subpopulation because a few blue GCs contaminate what are actually red subpopulation GCs in our defined green sample.

 As well as a possible trend of GC size with colour we also see a trend of GC size with galactocentric radius. In Figure \ref{fig:sizerad} a trend of increasing GC candidate size with increasing galactocentric radius is visible. %Median size values for each subpopulation are calculated in galactocentric radial bins, confirming that the trend is present at all colours. 
 Fitting a two parameter power law to the GC candidates brighter than $z=23.4$ we find 
\begin{equation} 
 r_h(\mathrm{pc})=[1.00\pm 0.04]R(\mathrm{kpc})^{[0.49\pm 0.04]} \notag
\end{equation} 
All three subpopulations show a clear size increase with distance from the galaxy centre with a power law slope of $0.49\pm 0.04$ dex per dex. This measurement extends to $4.5 R_e$, one of only four similar measurements extending beyond $\sim 2 R_{e}$. \citet{Ha09a} measured a slope of $0.11$ on a composite sample of six massive galaxies, \citet{Go07} measured a slope of $0.05\pm 0.05$ for metal poor GCs and $0.26\pm 0.06$ for metal rich GCs in NGC 5128 and \citet{Sp06} measured a slope of $0.19\pm 0.03$ in NGC 4594. The slope we measure is steeper than any of the previous values and much closer to the value measured for our own Galaxy ($0.36\pm0.07$ for metal rich GCs, see \citealt{Go07}). We conclude that there is more variability, between galaxies, in the relationship of GC size with galactocentric radius than previously found. Shown in Figure \ref{fig:sizerad} are the GCs in all three high probability subpopulation samples and power law fits. We set the slope of the fits to $0.49$ and found normalisations of $1.096\pm0.042$, $0.994\pm0.059$ and $0.944\pm0.044$ for the blue, green and red samples respectively. This results agrees with analysis of GC subpopulation median sizes.

\subsubsection{Mass function}
\begin{figure}\centering
 \includegraphics[width=0.47\textwidth]{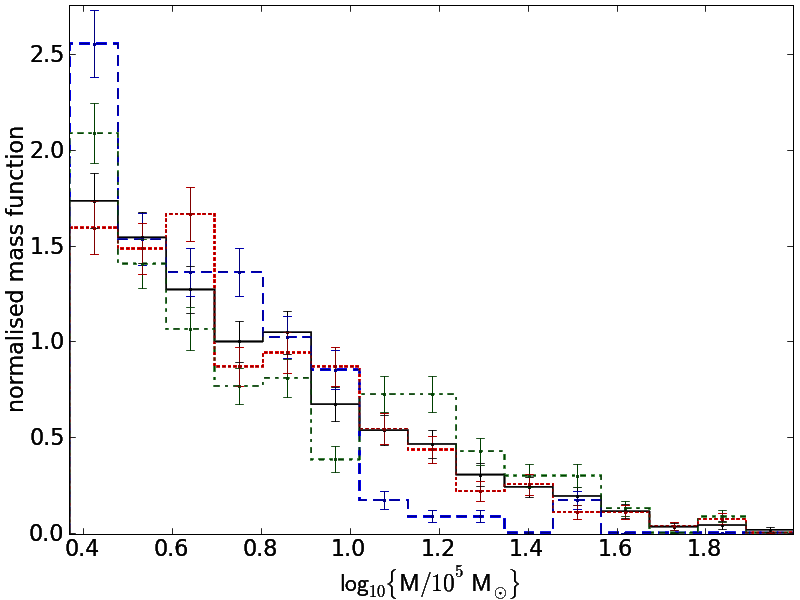}
 \caption{The normalised mass function for GC candidates brighter than $z=23.4$ from the ACS catalogue. All GCs (black) as well as the high probability subpopulation candidates (plotted in blue dashed, green dash-dotted and red dotted lines respectively). Each subpopulation has a significantly different mass distribution to the others.}
 \label{fig:massfn}
\end{figure}
 We compare the mass function of the high probability subpopulation GC candidates (as defined in Section 6.5.2). The mass for each GC candidate was derived using the colour dependent values for the $z$ filter mass-to-light ($\mathrm{M/L}_z$) ratio in Table 5 of \citet{VCS12}. They used PEGASE 2.0 with a modified Salpeter initial mass function from \cite{Ke83} and an age of 13 Gyr for all GCs. We used magnitudes brighter than $z=23.4$ to calculate the mass functions (plotted in Figure \ref{fig:massfn}). We ran the Kolmogorov-Smirnov (KS) test on each pair with a common mass range of $2\times10^{5}\mathrm{M}_{\odot}<\mathrm{M}<10^{7}\mathrm{M}_{\odot}$. The KS test gives a probability of 98.3 per cent that the blue and red subpopulations are drawn from different distributions and gives a probability of more than 99.9 (99.4) per cent that the green subpopulation is drawn from a different distribution than the blue (red) subpopulation. The discrepancy between the three subpopulations is greatest around $1\times10^{6}\mathrm{M}_{\odot}$ in the cumulative mass distribution the KS test uses for its analysis. %$10^{6.03}\mathrm{M}_{\odot}$ %or $1.0716\times10^{6}$

Analysis of the mass function of GC subpopulations indicates that the green subpopulation is distinct from both blue and red subpopulations, containing a significantly larger percentage of objects between $10^{6}\mathrm{M}_{\odot}$ and $10^{7}\mathrm{M}_{\odot}$ than either blue or red subpopulations, as can be clearly seen in Figure \ref{fig:massfn}. This overdensity of green subpopulation objects can also be seen in Figure \ref{fig:cmda} around $g-z \sim 1.1$ and $22>z>19$. It is possible that UCDs or dE nuclei contaminate the bright, mid-colour range \citep{St06} to produce this feature.
 
\subsubsection{Azimuthal properties}
\begin{figure}\centering
 \includegraphics[width=0.47\textwidth]{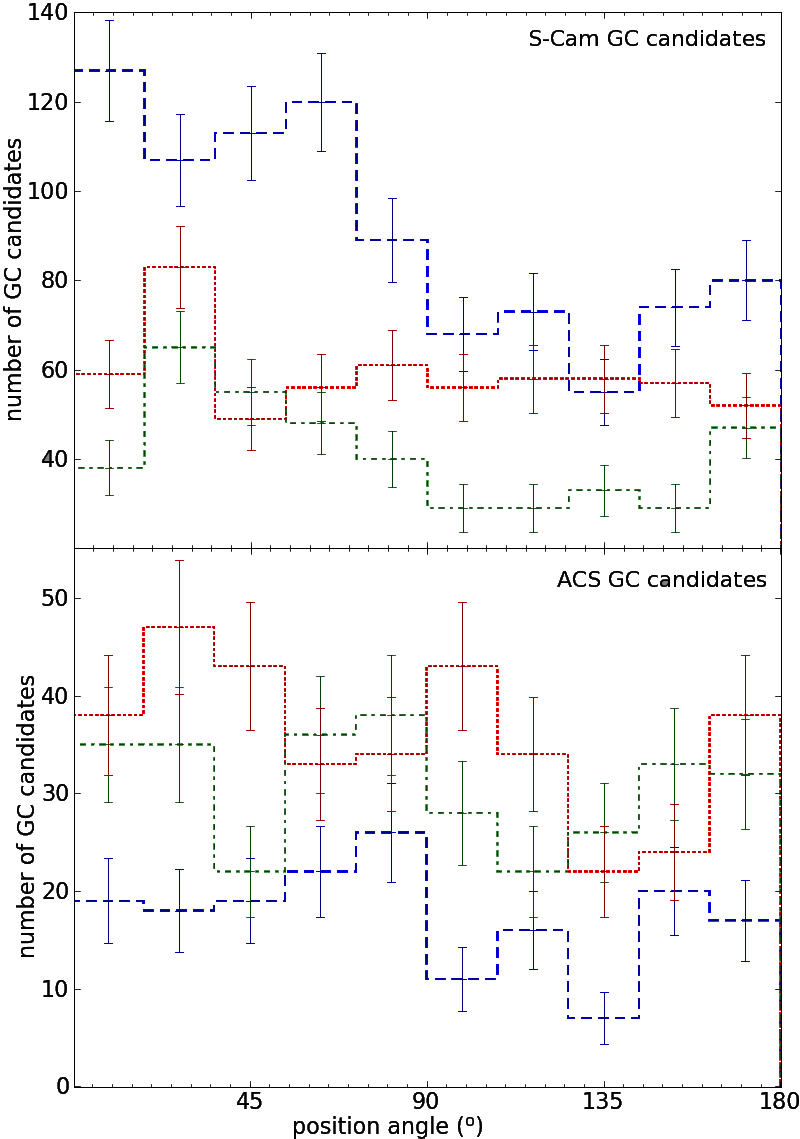}
 \caption{Histograms of the azimuthal distributions of high probability blue (dashed), green (dash-dotted) and red (dotted) GC subpopulations. The top panel shows the position angles for the S-Cam GC candidates out to $11$ arcmin from the galaxy centre and the bottom panel shows the ACS GC candidates out to $3.38$ arcmin from the galaxy centre. The position angle of the galaxy is $\sim 42^{\circ}$ and both blue and green subpopulations show GC overdensity between $0^{\circ}$ and $90^{\circ}$ in the S-Cam catalogue.}
 \label{fig:pacol}
\end{figure}
In Figure \ref{fig:pacol} we show the azimuthal distribution of high probability colour subpopulation GC candidates (as defined in Section 6.5.2) for S-Cam and ACS. We restrict the samples to radii where we have complete azimuthal coverage so as not to bias the samples in any direction. %There is no significant contamination from any nearby large galaxy and because NGC 4365 is centred in both sets of imaging we can extent the analysis to large radii without needing to make any corrections for non-uniform azimuthal coverage. 
The position angle data is folded at $180^{\circ}$ to improve number statistics. The bottom panel of Figure \ref{fig:pacol} shows that there is no position angle signal within errors for any of the three subpopulations in the ACS sample. The S-Cam azimuthal distribution in the top panel does show some structure; the high probability blue GC candidates are more likely to have position angles between $0^{\circ}$ and $90^{\circ}$, which is consistent with the blue GC subpopulation being elongated along the major axis of NGC 4365 (the galaxy has a position angle of $\sim 42^{\circ}$, see Section 5). The red distribution shows an almost flat azimuthal distribution (indicating an almost circular GC system) and the green distribution shows a very shallow sinusoidal distribution with a peak between $0^{\circ}$ and $90^{\circ}$.  %and this analysis would require much larger numbers of GC candidates to be able to discern any possible differences in the azimuthal structure of the green and red GC subpopulations. 
It is clear that the blue GC system is very elongated along the position angle of the galaxy and likely that both green and red subpopulations also have a similar position angle to the galaxy starlight. We obtain estimates of the ellipticity of the blue, green and red subpopulations by fixing the position angle (to $42^{\circ}$) as well as the power law exponent and background value as described in Section 6.1.1, Equation 4 and following the same fitting procedure. We find that the estimated ellipticity for the blue and green subpopulations are the same within errors and also consistent with the ellipticity measured for the whole GC sample in Section 6.1.1. The estimated ellipticity of the red subpopulation is much smaller than the blue and green values. It is consistent with zero and also the ellipticity measured for the galaxy light (see Table \ref{tab:ellip}). % with the exception of a strong peak in the red distribution at $25^{\circ}$.
%Combining Subaru/S-Cam and HST/ACS high probability colour subpopulation candidates allows a robust determination of the shape and position angle of the GC subpopulations. Because both the S-cam and ACS imaging of NGC 4365 that we use is centred on the galaxy, roughly radially symmetric and free of contamination from nearby large galaxies (see Figure \ref{fig:footp}) a simple combination of the samples is valid. Figure \ref{fig:pacol} shows histograms of the position angle of GC candidates that have a high probability of belonging to the blue, green or red subpopulations and while both green and red subpopulations have a relatively flat position angle distribution (indicating circular subpopulation shape) the blue subpopulation has a clear peak at $50^{\circ}$. 

We also split each subpopulation into an inner and outer radial bin with equal numbers in each and refit for ellipticity but did not detect radial variation in ellipticity for any of the three subpopulations.
\begin{table}\centering
 \begin{tabular}{cccc}
 \hline
\textbf{Galaxy Light} & \textbf{Blue} & \textbf{Green}  & \textbf{Red} \\ \hline \hline
$ 0.25\pm0.03$ & $0.66\pm0.06$ & $0.55\pm0.07$ & $0.16\pm0.25$ \\ \hline
 \end{tabular}
 \caption{Estimated ellipticity for blue, green and red GC subpopulations. The position angle of each GC subpopulation is fixed to be the same as that of the galaxy light ($42^{\circ}$).}
 \label{tab:ellip}
\end{table}

The agreement in ellipticity between the galaxy light and the red GC subpopulation is consistent with literature findings that red GC subpopulations generally follow the properties of galaxy field stars more closely than blue GC subpopulations. The blue GC subpopulation dominates in the outer regions of the galaxy and also dominates the measurement of the ellipticity of the GC system. It can be seen qualitatively in Figure \ref{fig:dens2d} and quantitatively from the results in Table \ref{tab:ellip} that the blue subpopulation is significantly more elliptical than the galaxy field stars. %It is puzzling that the green subpopulation has an ellipticity estimate that is consistent with the blue subpopulation when we expect the green subpopulation to be either a subset of the red subpopulation or a distinct group. %It is possible that the distinct green subpopulation is only found in the very central regions, that all GCs beyond $2$ arcmin are either blue or red contaminating objects and this ellipticity measurement is dominated by those outer blue contaminating objects.

\begin{figure}\centering
 \includegraphics[width=0.47\textwidth]{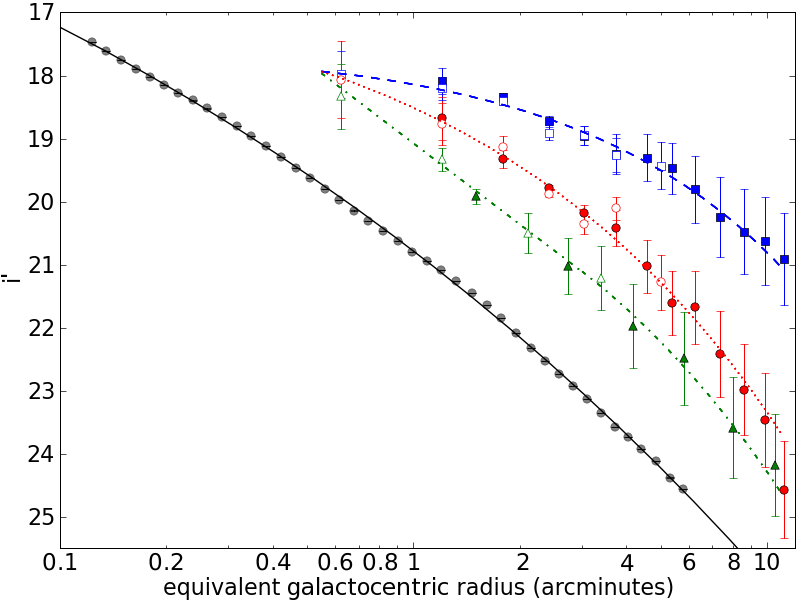}
 \caption{The radial surface brightness profile of the galaxy light from the S-Cam $i'$ filter compared with the surface density profile of the GC subpopulations. The GC subpopulations are plotted as blue squares, green triangles and red circles. The grey points show the surface brightness measurements and the black line shows the S\'{e}rsic profile fit to the galaxy light (see Section 5.1). All parameters are plotted against equivalent galactocentric radius, calculated as the geometric mean of the semi-major and semi-minor radii. The green and red GC surface density profiles are similar to the galaxy light surface brightness profile.}
 \label{fig:galsb}
\end{figure}
\subsection{Comparison with galaxy surface brightness}
The galaxy surface brightness profile from the $i'$ filter photometry is compared with the GC surface density profiles in Figure \ref{fig:galsb}. We compare the galaxy surface brightness profile with $2.5$ times the logarithm of the GC surface density profile and arbitrarily scaled for ease of viewing, i.e.\
\begin{equation}
2.5\mathrm{log}[P(R)-bg]
\end{equation}
where $P(R)$ is the GC radial surface density profile and $bg$ is the determined background value. Visually, the shape of the blue GC surface density profile is very different to that of the galaxy surface brightness whereas both red and green GC surface density profiles have shapes similar to the galaxy light. Both red and green density profiles are slightly steeper than the galaxy light profile beyond $\sim 5$ arcmin. % and while the green density profile is also steeper interior to $\sim 1$ arcmin, the red density profile is shallower than the galaxy light profile interior to $\sim 1$ arcmin.
 
\section{Discussion}
In previous work on NGC 4365, evidence of an additional subpopulation of GCs at intermediate (green) colours was based on small numbers and restricted to the central regions \citep{Pu02,La03,He04,Br05,Ku05,La05}. The large sample of GCs in this work, covering an extended radial range, allows us to revisit these claims in the optical wavelength range. %This work is restricted to the optical wavelength range and we note that features of the GC system colour distribution may change for a different intervals. 

We applied several independent statistical tests to the colour distribution, i.e.\ Chi-Squared minimization ($\chi^2$), the Kolmogorov-Smirnov (KS) test and Kaye's Mixture Model (KMM) algorithm. When the entire radial range of our imaging is used our statistical tests cannot conclusively rule out a bimodal interpretation, partly because GC subpopulations do not have strictly Gaussian colour distributions. %The reduced $\chi^2$ value decreases from the bimodal to trimodal case but the third mode is not central in colour and the KS test rejects both bimodal and trimodal Gaussians. The KMM test statistic prefers trimodality over bimodality but this version of the test has not been applied to the colour distributions of other GC systems (see Appendix A). %When the radial range is restricted, the $\chi^2$ minimization indicates more strongly that trimodality is preferred.                       %%%They are the KMM, KS and Chi Squared minimization tests. This photometric analysis of NGC 4365's GC system has been largely concerned with the third GC subpopulation. Not all recent publications are in agreement that it is necessary to attribute more than the usual two GC subpopulations to this GC system and          Deeper, wide-field imaging has increased the number of observed GC candidates in many GC systems and trends of GC colour with size, luminosity and galactocentric radius have been discovered. We can see the impact of these trends on the GC colour distribution and notice that without restrictions on luminosity, galactocentric radius and GC size it is no longer accurate to describe GC colour distributions in terms of symmetric Gaussians. The statistical tests used here to test for trimodality cannot accommodate asymmetric Gaussians and cannot conclusively rule out either a symmetric bimodal or trimodal interpretation for the GC system of NGC 4365. These tests have not given clear answers about the reality of a third GC subpopulation. In all cases results from tests were inconclusive about whether a split into three subpopulations was preferred over a split into two subpopulations. Thus, from a purely statistical analysis there is no reason to favour a trimodal description of NGC 4365's GC colour distribution over a bimodal one. In the case where the red GC subpopulation has a fairly strong colour gradient with galactocentric radius and the green subpopulation is limited to the very centre of the galaxy (within $1'$ or $2'$) this confusion in the colour distribution (collapsed in galactocentric radius) is easily understood. We do find evidence of both a very centrally concentrated overdensity of intermediate colour GC candidates and a gradient in the red subpopulation to intermediate colours at larger galactocentric radii, indicating that statistical tests based purely on the colour distribution cannot disentangle the possible three GC subpopulations of NGC 4365. 

When examining GC subpopulation colours with radius we find significant negative colour/metallicity gradients in both the bimodal and trimodal cases. Due to its radial colour gradient, the red subpopulation has somewhat greener colours at large radii. This explains why statistical tests showed inconclusive results when the full radial extent of the imaging was used. However, this cannot explain the central green overdensity that motivates splitting NGC 4365's GCs into three subpopulations.

We compared the photometrically observable properties of blue, green and red GCs (measured here to contain 43, 17 and 40 per cent of  total GCs respectively). The high probability blue, green and red samples have colour ranges of $g-i<0.82$, $0.89<g-i<1.0$ and $g-i>1.1$ where definitions are based on kernel smoothing of the GC colour distribution. We find evidence for different properties between the subpopulations. %Specifically we find several differences between the green and red subpopulations. (radial and azimuthal distributions, sizes and mass functions)
The green GCs are intermediate in size to the blue and red subpopulations (clearly smaller than the blue GCs and marginally larger than the red ones) and have a significantly different mass function to both blue and red subpopulations. The size differences could be due to a continuous trend with colour rather than specific subpopulation properties \citep[see][]{Jo04} but the differences in mass functions suggest that the three subpopulations are indeed distinct. Green GCs also have a significantly steeper surface density profile than red GCs within $2$ arcmin ($13.4$ kpc) from the galaxy centre. However, the green and red subpopulations can be described by the same radial surface density profile, with different normalisations, beyond $2$ arcmin ($13.4$ kpc). % bringing to mind the possibility that beyond the central region the `green' GCs are actually mostly red GCs. 
The ellipticity of the green GC subpopulation is similar to the blue GC subpopulation and significantly more elongated than that of the red. We do not detect any variation in ellipticity of the subpopulations with galactocentric radius to 11 arcmin (74 kpc). % and the green GCs have the same elongation internal to and beyond $2$ arcmin ($13.4$ kpc), which seems to contradict this possibility. %However, the ellipticity measurement of the greens may be dominated by blue GCs in the outer parts.
These results lend support to the idea that blue, green and red GCs form three distinct subpopulations. %with the green GCs only situated in the very inner parts of the galaxy. All intermediate colour GCs beyond $\sim 2$ arcmin ($13.4$ kpc) are likely to be tails from blue and red subpopulation distributions. %Using the same subpopulation division described above we

With the same three subpopulation division we find a number of similarities between the properties of the red subpopulation and the galaxy light of NGC 4365. The $g-i$ radial colour gradient of the galaxy light and mean colour of the red GC subpopulation are virtually identical \citep[see also][]{Sp10,Fo01}. The ellipticity of the red subpopulation is consistent with that of the galaxy starlight and the surface density profile of the red subpopulation is closer to the starlight surface brightness profile than either blue or green subpopulations. These results confirm similar findings in the literature \citep{Br06} and support GC formation scenarios in which the red subpopulation is closely linked to the formation of the bulk of the galaxy field stars. \\

%This supports the idea that both red and green GCs should properly be called red GCs and the GC colour distribution of NGC 4365 is bimodal. %The possible green subpopulation of GCs in NGC 4365 is very centrally concentrated, has sizes slightly larger than the red subpopulation GC candidates and a greater percentage of objects with masses larger than $\sim10^6\mathrm{M}_{\odot}$ than either the blue or red subpopulations.
To understand the physical implications of optical colour trimodality we consider the relationship between optical colours, metallicities and ages.
%We assume that the colour distribution of the GCs is a proxy for the metallicity distribution of the GCs, where all GCs are old and the transformation from colour to metallicity is close to linear. 
There is considerable debate about whether the colour distribution of GC systems is a good proxy for the metallicity distribution, even if uniform old ages for GCs are assumed \citep[see discussions on the non-linearity of colour-metallicity transformations by][]{Y06,Ca07,He07}. %Using the \citet{Le10b} empirical relationship between GC colour and metallicity the blue, green and red colour peaks we adopt correspond to [Fe/H] values of -1.35, -0.72 and -0.20. 
It is also possible that one or more of the GC subpopulations have significantly younger ages caused by a merger \citep[see][]{Hi95,Sc98,Mo10} though there is little spectroscopic evidence of young ages for NGC 4365 GCs \citep{Br05}. %Our subpopulation split has 43, 17 and 40 per cent of GCs in the blue, green and red subpopulations respectively. %The colour metallicity transformation has been claimed to be non-linear for optical colours. Several papers discuss the possibility that a bimodal colour distribution could have an underlying unimodal metallicity distribution due to the non-linearity of the metallicity-colour transformation. It is possible that NGC 4365's GC system is not trimodal but bimodal or unimodal in metallicity but trimodal in colour. It is also possible that other giant elliptical galaxies have trimodal metallicity distributions but normal bimodal colour distributions. Only a handful of giant elliptical galaxies have spectroscopic metallicities for a good fraction of their GCs.

%If NGC 4365 does have three distinct GC subpopulations (measured here to contain 43, 17 and 40 per cent of GCs in the blue, green and red subpopulations respectively) %visible in optical colours and if the three subpopulations found in colour correspond to three metallicity subpopulations 
The green subpopulation might be a consequence of a unique evolutionary history of NGC 4365. The kinematically distinct core (KDC) in the stellar light might be a signal of this. This property is only found in galaxies the SAURON team \citep{dZ02} classify as slow rotators in the inner regions \citep{Em07}. The KDC is confined within $5$ arcsec ($0.56$ kpc) of the galaxy centre and the stellar population inside and outside the KDC are indistinguishable in age and metallicity. Inside $5$ arcsec the stars rotate around the minor axis and outside $5$ arcsec they rotate around the major axis \citep{Da01}. \citet{Da01} suggest that the KDC might have been formed in the merger of ``gas-rich fragments at high redshift" and state that they find no evidence of recent star formation in NGC 4365. The effective radius of the galaxy ($\sim 2$ arcmin or $\sim 13.4$ kpc) and the radial extent of the green subpopulation ($\sim 2$ arcmin) are much larger than the KDC. It is possible that the formation of the green GC subpopulation is linked to the formation of the KDC but since they are not spatially correlated, and other elliptical galaxies with KDCs show no clear indication of a green subpopulation, the connection is not clear.

The core must have been formed very early on because \citet{Da01} find the same $>12$ Gyr luminosity-weighted age for the KDC and the rest of NGC 4365, with no sharp changes in metallicity across the boundary of the KDC. If the formation of the green subpopulation is linked then we would expect the green subpopulation to also be very old. \citet{Br05} found evidence that the green subpopulation was indeed old but this analysis was based on a small number of GCs. We might also expect to find rotation for the green subpopulation to be offset by $\sim 90^{\circ}$ compared to the other GC subpopulations and the bulk of the starlight, if it is associated with the formation of the KDC. %KDCs are thought to form as a result of major mergers that ejects stars and gas, which later fall into the galaxy centre with a rotation signature different to the bulk of the galaxy.% (within 12 Gyr from today).

Another possible signature of the unique evolutionary history could be the significant misalignment between the galaxy's kinematic and photometric major axes. While KDCs are relatively common in elliptical galaxies, the minor axis rotation of the bulk of NGC 4365's starlight is relatively uncommon. Van den Bosch et al.\ (2008) use triaxial orbit based models to explain NGC 4365's apparent minor axis rotation and KDC. The combination of axisymmetry in the inner parts with triaxiality in the outer parts that \citet{vdB08} describe is consistent with the age and metallicity of the KDC but bears no obvious relation to the formation of a green GC subpopulation in NGC 4365. %Van den Bosch et al.\ (2008) use triaxial orbit based models to conclude that NGC 4365 appears to rotate about its major axis because the bulk of its starlight has triaxial rather than axisymmetric orbital structure, and it appears to have a kinematically distinct core because the inner regions are nearly axisymmetric. 
Alternatively, \citet{Ho10} show that a $\sim2.5$ Gyr old major merger remnant with $15 - 20$ per cent progenitor gas fraction has a kinematic signature remarkably similar to NGC 4365. The gas rich major merger that \citet{Ho10} describe is a tempting explanation for the presence of green GCs because all NGC 4365's anomalies would have one explanation. Given that the green GC subpopulation is likely old \citep{Br05} and there is no evidence for young stars in the KDC \citep{Da01} this would have to be a very early major merger. \\

The current galaxy and GC system formation scenarios that describe mechanisms for producing bimodality in GC systems could also explain GC system trimodality for a subset of galaxies under certain conditions.

In the multiphase collapse scenario GC systems are formed during two phases, the first phase produces metal poor GCs, is truncated by one of a variety of possible mechanisms and then later the second phase produces the metal rich GCs \citep{F97}. In this context a GC system that is trimodal in metallicity could simply be the result of three formation phases truncated twice (by the same or two different mechanisms). The truncation of the metal poor formation phase could be caused by a universal epoch of reionization but, for truncation to happen twice, at least one truncation mechanism has to be related to nearby galaxies or a process internal to the galaxy, like an active galactic nucleus. 

The major merger formation scenario \citep{Ze93}, where metal poor GCs are present in galaxies before mergers and metal rich GCs are formed during major mergers, could be consistent with some galaxies containing three GC subpopulations. If a galaxy underwent a very early gaseous major merger and later another major merger it may have two subpopulations more metal rich than the metal poor GC subpopulation. Depending on the age and metallicity difference it could appear trimodal in colour.

The accretion model for GC system formation \citep{C98} found more than two modes in some of their simulated GC systems before a third GC subpopulation was considered for NGC 4365 from observations. \citet{C98} mention that more than two peaks are present in some of their simulations when a very steep luminosity function (Schechter function $\alpha=-1.8$) is used. They find that the peak of the metal poor population correlates with the slope of the luminosity function and that for very steep slopes the metal poor peak is more metal poor. It is conceivable that the merger histories of some giant elliptical galaxies will therefore show the presence of a third, old, intermediate metallicity subpopulation.
%WRONGThe accretion model for GC system formation does not have a natural origin for a subpopulation of intermediate metallicity GCs found in some galaxies. In this scenario the metal rich GCs are formed in the deep potential well of large galaxies and the metal poor GCs are accreted from dwarf galaxies where they formed in shallow potential wells. If there is a significant population of intermediate metallicity GCs in the universe, formed in medium depth potential wells, they should be found as an intermediate metallicity population in most giant elliptical galaxies today. The accretion scenario does not well explain a few anomalous GC systems with intermediate metallicity subpopulations.

The questions that remain include distinguishing which of these formation scenarios best explain the GC system of NGC 4365 and determining whether the trimodality in the colour distribution of the GC system is reflected in the metallicity distribution. The method in \citet{Fos10} for determining metallicity from the Calcium triplet indices could provide a large sample of metallicity measurements over a wide field of view to assess whether the metallicities of the three colour subpopulations we define here are distinct. The spectroscopic analysis of a large number of GCs could also be used to build a picture of the kinematics of the GC system of NGC 4365 and its subpopulations. An understanding of NGC 4365's GC system kinematics could distinguish whether two or three subpopulations were all formed in situ, whether the more metal rich populations were accreted or whether the subpopulations are the result of several major mergers.

\section{Summary and Conclusions}
Combining the photometric depth, size information and resolution of HST/ACS data with the spatial extent and three filter imaging of the Subaru/S-Cam data we can achieve a uniquely detailed, and unmatched spatially extensive, analysis of NGC 4365's GC system. The GC system extends beyond 134 kpc from the galaxy centre to $\>9.5R_e$. The blue GC subpopulation has not yet reached the background level at the very edges of our Subaru/S-Cam imaging. We place a lower limit on the total number of GCs  to be $6450\pm110$. 

We find further evidence to support the existence of a distinct third subpopulation at intermediate colours in the GC system of NGC 4365. %We however cannot rule out the possibility of adequately describing this GC system in terms of two subpopulations. %with size and colour trends within the red subpopulation is as good a description of the GC system of NGC 4365. 
We also find a trend of increasing GC size with galactocentric radius and negative gradients in the colour/metallicity of both blue and red subpopulations with galactocentric radius. The blue subpopulation shows evidence for a blue tilt. Comparing the GC system with the galaxy light we find that the red subpopulation has a similar colour, radial colour gradient, ellipticity and radial surface density slope to the galaxy light.

Based on various measured properties we find it most likely that the green subpopulation is distinct from the blue and red subpopulations. Consequently, it is very likely that NGC 4365 has had a unique evolutionary history causing the existence of a third GC subpopulation in this giant elliptical, where most similar galaxies only have two. The photometric properties of the blue and red GC subpopulations are consistent with the properties found for those of other giant elliptical GC systems, therefore any formation scenario for the green subpopulation must leave the predicted properties of the other two subpopulations relatively unchanged.

Future analysis of spectroscopy already obtained for this system will help determine the origin of the green subpopulation of NGC 4365's GC system.

\section{Acknowledgements}
We thank the reviewer, A.\ Graham, C.\ Foster, M.\ Owers, A. Romanowsky and A.\ Chies-Santos for insightful discussions. We also thank M.\ Smith, V.\ Pota, C.\ Usher and S.\ Kartha for support during the preparation of this manuscript. Based in part on data collected at Subaru Telescope, which is operated by the National Astronomical Observatory of Japan. Based in part on observations made with the NASA/ESA Hubble Space Telescope, and obtained from the Hubble Legacy Archive, which is a collaboration between the Space Telescope Science Institute (STScI/NASA), the Space Telescope European Coordinating Facility (ST-ECF/ESA) and the Canadian Astronomy Data Centre (CADC/NRC/CSA).

%
% References
%
\bibliographystyle{mn2e}
\bibliography{ref}

\begin{appendix}
\section{Statistical tests for trimodality}
%\textit{I am aiming to get to expanding this by the end of the week, it needs a few tables (KMM results, KS results, $\chi^2$ results) and plots ($\chi^2$ Gaussians with Kernel smoothed data)}
\subsection{Kaye's Mixture Model algorithm}
\begin{table*}\centering
 \begin{tabular}{@{}cc@{}|c@{}|ccccccccccccc@{}} \hline
 \multicolumn{16}{|c|}{\textbf{Comparing galaxies with Subaru/S-Cam photometry and $g'-i'$ colours where $i'\leqslant23$}} \\ \hline \hline
 \multirow{2}{*}{\textbf{Galaxy}} & \multicolumn{3}{|c|}{\textbf{Unimodal}} & \textbf{p} & \multicolumn{3}{|c|}{\textbf{Bimodal}} & \textbf{p} & \multicolumn{3}{|c|}{\textbf{Trimodal}} & \textbf{p} & \multicolumn{3}{|c|}{\textbf{Quadrumodal}}\\
 & $\mu$ & $\sigma$ & $n$ & (2 over 1) & $\mu$ & $\sigma$ & $n$ & (3 over 2) & $\mu$ & $\sigma$ & $n$ & (4 over 3) & $\mu$ & $\sigma$ & $n$\\ \hline \hline
 \multirow{4}{*}{NGC 4365} & 0.95 & 0.16 & 1679 & \multirow{4}{*}{0.000} & 0.79 & 0.05 & 642 & \multirow{4}{*}{0.003} & 0.78 & 0.04 & 600 & \multirow{4}{*}{0.252} & 0.78 & 0.05 & 642\\
 & & & & & 1.05 & 0.12 & 1037 & & 0.96 & 0.09 & 567 & & 0.96 & 0.09 & 298\\
 & & & & & & & & & 1.13 & 0.09 & 512 & & 1.01 & 0.1 & 302\\
 & & & & & & & & & & & & & 1.15 & 0.08 & 437\\ \hline
 \multirow{4}{*}{NGC 1407} & 0.92 & 0.16 & 1604 & \multirow{4}{*}{0.000} & 0.79 & 0.06 & 784 & \multirow{4}{*}{0.000} & 0.76 & 0.04 & 616 & \multirow{4}{*}{-} & 0.77 & 0.04 & 696\\
 & & & & & 1.06 & 0.09 & 820 & & 0.93 & 0.11 & 562 & & 0.91 & 0.12 & 76\\
 & & & & & & & & & 1.13 & 0.06 & 426 & & 0.98 & 0.1 & 415\\
 & & & & & & & & & & & & & 1.14 & 0.06 & 417\\ \hline
 \multicolumn{16}{|c|}{\textbf{Comparing galaxies with HST/ACS photometry and $g-z$ colours}} \\ \hline \hline
 \multirow{3}{*}{NGC 4365} & 1.12 & 0.22 & 1658 & \multirow{4}{*}{0.000} & 0.89 & 0.08 & 605 & \multirow{4}{*}{0.003} & 0.88 & 0.08 & 550 & \multirow{4}{*}{0.596} & 0.88 & 0.08 & 584\\
 & & & & & 1.25 & 0.15 & 1053 & & 1.12 & 0.12 & 488 & & 1.11 & 0.12 & 203\\
 $z\leqslant24$ & & & & & & & & & 1.13 & 0.12 & 620 & & 1.19 & 0.14 & 348\\
 all point.\ & & & & & & & & & & & & & 1.36 & 0.11 & 523\\ \hline
 \multirow{3}{*}{NGC 4365} & 1.19 & 0.24 & 905 & \multirow{4}{*}{0.000} & 0.94 & 0.13 & 318 & \multirow{4}{*}{0.082} & 0.92 & 0.07 & 120 & \multirow{4}{*}{0.915} & 0.91 & 0.15 & 45\\
 & & & & & 1.33 & 0.17 & 587 & & 0.98 & 0.17 & 192 & & 0.92 & 0.07 & 204\\
 $z\leqslant24$ & & & & & & & & & 1.33 & 0.17 & 593 & & 1.08 & 0.19 & 61\\
 cent.\ point.\ & & & & & & & & & & & & & 1.34 & 0.17 & 595\\ \hline
 \multirow{3}{*}{M60} & 1.24 & 0.26 & 795 & \multirow{4}{*}{0.000} & 0.98 & 0.13 & 320 & \multirow{4}{*}{0.328} & 0.85 & 0.11 & 51 & \multirow{4}{*}{-} & 0.83 & 0.10 & 43\\
 & & & & & 1.43 & 0.15 & 475 & & 1.00 & 0.09 & 255 & & 1.00 & 0.09 & 272\\
 $z\leqslant24$ & & & & & & & & & 1.41 & 0.16 & 489 & & 1.29 & 0.08 & 147\\
 cent.\ point.\ & & & & & & & & & & & & & 1.49 & 0.13 & 333\\ \hline
 \end{tabular}
 \caption{Results from the extended KMM test for testing the significance with which $m+1$ modes are preferred over $m$ modes in colour. For each mode the mean value ($\mu$), width ($\sigma$) and number of objects assigned ($n$) are shown and between each modality the `p' statistic is shown.}
 \label{tab:kmmvals}
\end{table*}
The most common tool to determine the statistical significance of GC bimodality is the Kaye's Mixture Model (KMM) code \citep{KMM}, designed to calculate the `p' statistic for the significance of a homoscedastic bimodal Gaussian distribution being preferred over a unimodal Gaussian distribution. KMM is also capable of determining the significance of heteroscedastic multimodal distributions being preferred over a unimodal Gaussian distribution but the `p' statistic in these cases is an approximation. Applied to our GC colour distribution for NGC 4365 both bimodal and trimodal distributions result in a `p' statistic less than $10^{-4}$ (a probability larger than 99.99 per cent that the multimodal distributions are preferred over the unimodal distribution). The traditional KMM code is however not capable of determining the probability that a trimodal distribution is preferred over a bimodal distribution. 

Using an extension to the traditional KMM code, written by Dr Matthew Owers \citep{Ow11}, we were able to assess a probability for three Gaussian modes over two Gaussian modes bearing in mind that the code was untested for application to GC colour distributions. We attempted to test the code at the same time as using it for our analysis by including checks on how the code performed with varying GC input numbers, analysis of other galaxies' GC distributions and redundant calculations for the probability of four modes over three modes and bimodality over unimodality. The code showed variation of results with input GC number and gave results for other galaxies inconsistent with results in literature. Comparing results for NGC 4365 with NGC 1407 (data from Spitler et al.\ 2011 in prep. also used in \citealt{Ro09,Fos10,Fo11}) the extended KMM test finds that both are trimodal (NGC 4365 with a probability of 99.7 per cent and NGC 1407 with a probability larger than 99.9 per cent) but there is little evidence in the literature for NGC 1407 being trimodal. When the number of GCs is halved for NGC 4365 \citep[data from][]{VCS9} the probability of trimodality decreases below 95 per cent (to 91.8 per cent), even though the smaller sample is more central and we expect trimodality to be more prominent there.

%We decided to try alternative methods for determining the significance of trimodality over bimodality rather than spending time investigating the conditions under which the code results could be trusted, even though it did give some indication of a statistically significant third subpopulation with colours intermediate to the blue and red subpopulations.

\subsection{The Kolmogorov-Smirnov test}
\begin{table}\centering
\begin{tabular}{@{}ccccc@{}}\hline
\textbf{VCC number} & \textbf{1226} & \textbf{1316} & \textbf{731} & \textbf{1535} \\ 
\textbf{Other name} & \textbf{M49} & \textbf{M87} & \textbf{NGC 4365} & \textbf{NGC 4526} \\ \hline \hline
\textbf{M49} & 0.79 & 0.86 & $1.6\times10^{-7}$ & $1.4\times10^{-3}$ \\
\textbf{M87} & - & 0.45 & $4.6\times10^{-9}$ & $6.0\times10^{-4}$ \\
\textbf{NGC 4365} & - & - & - & $9.2\times10^{-3}$ \\ \hline
\end{tabular}
\caption{KS test results comparing galaxies in the ACS Virgo Cluster Survey. The `p' statistic shown is an inverse measure of the probability that GCs in the different galaxies are drawn from different underlying colour distributions. Only M49 and M87 are inconsistent with being drawn from different colour distributions and hence could be said to have the same underlying colour distribution. Other galaxy comparisons show a probability greater than 99 per cent of being drawn from different underlying colour distributions.}
\label{tab:ksgal}
\end{table}
The Kolmogorov-Smirnov (KS) test shows clearly that most giant elliptical galaxies have GC colour distributions that are significantly different from each other (subpopulation peak colours and widths vary between galaxies, see Table \ref{tab:ksgal}) \citep[using GC catalogues from][]{VCS9}. A sample of galaxies with reasonably large GC numbers were chosen to compare to each other and most galaxies show a probability of larger than 99 per cent of being drawn from a colour distribution different to any other galaxy in the sample. Two very large galaxies (M87 and M49) are consistent with being drawn from the same underlying colour distribution. It is clear that each galaxy's GC system is formed independently of any other GC system but that similar galaxy properties can result in similar GC systems.

\begin{table}\centering
\begin{tabular}{lrc}\hline
Simulated & \multicolumn{2}{c}{Actual colour distributions} \\
distributions & \textbf{ACS} & \textbf{S-Cam} \\ \hline \hline 
\textbf{Bimodal} & $10^{-95}$ & $10^{-256}$ \\
\textbf{Trimodal} & $10^{-99}$ & $10^{-262}$ \\ \hline
\end{tabular}
\caption{KS test results comparing simulated bimodal and trimodal distributions with NGC 4365's colour distribution. The `p' statistic shown is an inverse measure of the probability that the simulated distributions are drawn from different colour distributions to the actual ones.}
\label{tab:ksbitri}
\end{table}
We also used this test to compare colour distributions simulated from bimodal and trimodal descriptions of the data (as output by KMM) with the actual GC colour distribution and found that the observed GC distribution is significantly different from both simulated distributions. Table \ref{tab:ksbitri} shows that for all comparisons the KS test shows probabilities of almost 100 per cent that the simulated and actual distributions are significantly different. Neither simulated distribution is a good enough description of the observed distribution, likely because they cannot take into account the skewness and peakiness (kurtosis) of the observed distributions. The colour modes in GC systems are not perfectly Gaussian and therefore this method of analysis cannot distinguish between bimodal and trimodal Gaussian descriptions of the colour distribution to determine which is more likely.

\subsection{Chi-Squared minimization}
Comparing the bimodal and trimodal results for a reduced Chi-Squared ($\chi^2$) minimization is dependent on binning the data in colour. We have chosen to do this binning via an Epanechnikov kernel smoothing of the data rather than an histogram. With different colour bin sizes the statistical results will also be different. We have chosen bin sizes for the Epanechnikov kernel that are equal to the colour errors in the data.
 
Using the entire radial range of our data we can achieve the best number statistics but the results are not conclusive. For the ACS GCs, where we have better colour resolution, the best-fitting trimodal Gaussian distribution has a slightly smaller reduced $\chi^2$ value than the best-fitting bimodal distribution but the best fit third mode is very close to blue modes and contains many fewer objects. This `third mode' is not best described as a third GC subpopulation but as an indication that the two modes of the underlying GC distribution are not pure Gaussian but probably Gaussians with both skewness and kurtosis. We do see a significantly smaller reduced $\chi^2$ value and a central third mode with a large proportion of the GCs in the S-Cam catalogue.
\begin{figure}\centering
 \includegraphics[width=0.47\textwidth]{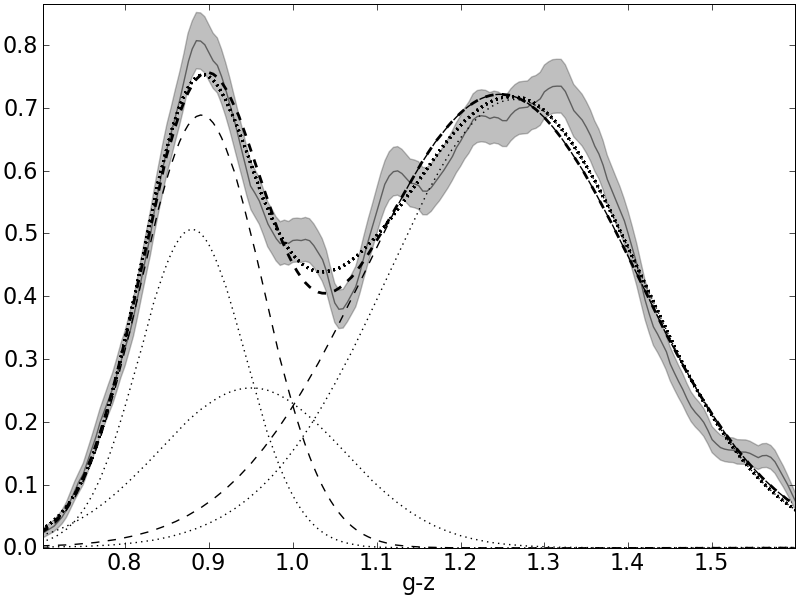}
 \caption{Results of bimodal and trimodal fits to the Epanechnikov kernel smoothing of the HST/ACS GC colour distribution. The distribution is plotted in grey, showing poissonian errors and the individual Gaussians as well as the summation of all three are overplotted for both the bimodal (dashed lines) and trimodal (dotted lines) fits. The reduced $\chi^2$ value is 1.40 for the bimodal case and 1.26 for the trimodal case.}
 \label{fig:achi}
 \end{figure}
 \begin{figure}\centering
 \includegraphics[width=0.47\textwidth]{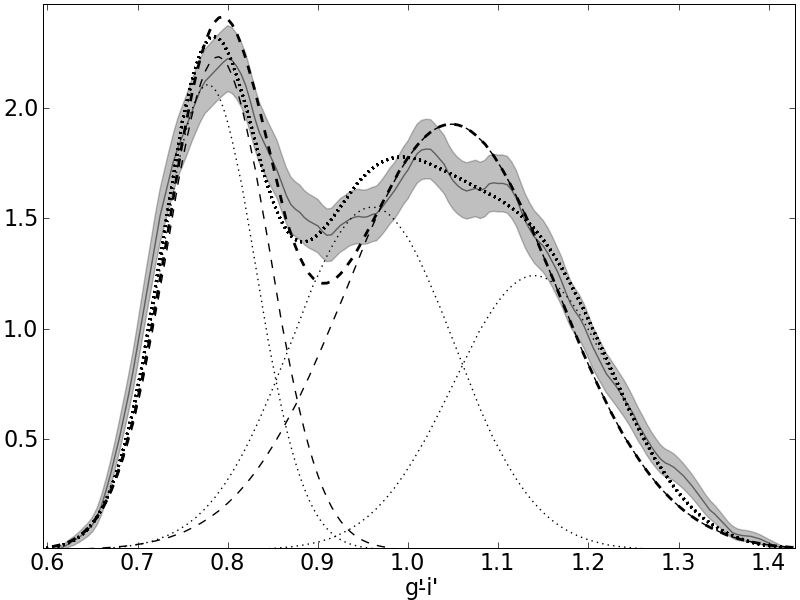}
 \caption{Results of bimodal and trimodal fits to the Epanechnikov kernel smoothing of the Subaru/S-Cam GC colour distribution. The distribution is plotted in grey, showing poissonian errors and the individual Gaussians as well as the summation of all three are overplotted for both the bimodal (dashed lines) and trimodal (dotted lines) fits. The reduced $\chi^2$ value is 2.57 for the bimodal case and 1.07 for the trimodal case.}
 \label{fig:schi}
 \end{figure}

%\bigskip \bigskip \bigskip \bigskip \noindent For various reasons none of the statistical tests employed conclusively rule out a bimodal interpretation of NGC 4365's colour distribution. The extended KMM and $\chi^2$ minimization marginally favour trimodality over bimodality in certain cases. 

\section{Surface density data}
\begin{table*}\centering
\begin{tabular}{cclrccclrcl}\hline
\multicolumn{11}{c}{\textbf{GC candidate radial surface density obtained from HST/ACS photometry}} \\ \hline \hline
%\multicolumn{3}{l}{\textbf{Total GC system}} & \multicolumn{5}{c}{\textbf{Blue and Red GC subpopulations}} & \multicolumn{3}{c}{\textbf{Green GC subpopulation}} \\
\textbf{r}$_{gal}$ & \textbf{Total} & \textbf{err} & \textbf{r}$_{gal}$ & \textbf{Blue} & \textbf{err}$_{B}$ & \textbf{Red} & \textbf{err}$_{R}$ & \textbf{r}$_{gal}$ & \textbf{Green} & \textbf{err}$_{G}$ \\ 
(arcmin) & \multicolumn{2}{c}{(arcmin$^{-2}$)} & (arcmin) & \multicolumn{2}{c}{(arcmin$^{-2}$)} & \multicolumn{2}{c}{(arcmin$^{-2}$)} & (arcmin) & \multicolumn{2}{c}{(arcmin$^{-2}$)} \\ \hline \hline
\textbf{0.20} & 119 & 30 & \textbf{0.62} & 12.5 & 2.3 & 38.0 & 4.1 & \textbf{0.63} & 24.6 & 3.3 \\
\textbf{0.32} & 125 & 20 & \textbf{1.2} & 10.2 & 1.5 & 20.0 & 2.1 & \textbf{1.2} & 9.9 & 1.5 \\
\textbf{0.50} & 81 & 11 & \textbf{1.8} & 8.5 & 1.1 & 14.5 & 1.5 & \textbf{2.1} & 3.51 & 0.46 \\
\textbf{0.75} & 66.5 & 6.8 & \textbf{2.4} & 5.48 & 0.74 & 7.57 & 0.86 & \textbf{3.4} & 1.92 & 0.27 \\
\textbf{1.10} & 43.4 & 4.0 & \textbf{3.0} & 5.27 & 0.69 & 5.06 & 0.67 & & &  \\
\textbf{1.50} & 31.6 & 2.9 & \textbf{3.8} & 4.03 & 0.52 & 6.22 & 0.64 & & &  \\
\textbf{1.90} & 26.6 & 2.4 & \textbf{5.0} & 3.47 & 0.40 & 2.44 & 0.33 & & &  \\
\textbf{2.30} & 14.9 & 1.6 & & & & & & & & \\
\textbf{2.75} & 15.9 & 1.4 & & & & & & & & \\
\textbf{3.15} & 13.6 & 1.5 & & & & & & & & \\ \hline
\multicolumn{11}{c}{\textbf{GC candidate radial surface density obtained from Subaru/S-Cam photometry}} \\ \hline \hline 
%$r_{gal}$ & $\rho$ & $\epsilon_{\rho}$ & $r_{gal}$ & $\rho_{B}$ & $\epsilon_{\rho_{B}}$ & $\rho_{R}$ & $\epsilon_{\rho_{R}}$ & $r_{gal}$ & $\rho_{G}$ & $\epsilon_{\rho_{G}}$ \\ \hline \hline
\textbf{1.10} & 43.4 & 4.0 & \textbf{1.2} & 11.4 & 1.6 & 21.8 & 2.2 & \textbf{1.5} & 5.86 & 0.73 \\
\textbf{1.50} & 32.0 & 2.9 & \textbf{1.8} & 9.0 & 1.2 & 12.3 & 1.4 & \textbf{2.7} & 2.25 & 0.32 \\
\textbf{1.90} & 24.9 & 2.3 & \textbf{2.4} & 6.39 & 0.79 & 8.21 & 0.90 & \textbf{4.2} & 1.05 & 0.16 \\
\textbf{2.30} & 16.4 & 1.7 & \textbf{3.0} & 5.23 & 0.69 & 5.85 & 0.73 & \textbf{5.8} & 0.74 & 0.11 \\
\textbf{2.75} & 14.9 & 1.3 & \textbf{3.8} & 4.06 & 0.45 & 4.79 & 0.49 & \textbf{8.0} & 0.399 & 0.056 \\
\textbf{3.25} & 11.8 & 1.1 & \textbf{4.5} & 3.88 & 0.40 & 2.94 & 0.35 & \textbf{10.5} & 0.318 & 0.044 \\
\textbf{3.75} & 10.70 & 0.95 & \textbf{5.4} & 3.34 & 0.36 & 1.93 & 0.28 & \textbf{13} & 0.328 & 0.038 \\
\textbf{4.25} & 8.54 & 0.80 & \textbf{6.2} & 2.52 & 0.25 & 1.85 & 0.22 & \textbf{16} & 0.174 & 0.024 \\
\textbf{4.75} & 7.57 & 0.71 & \textbf{7.4} & 1.77 & 0.17 & 1.18 & 0.14 & & &  \\
\textbf{5.25} & 5.88 & 0.60 & \textbf{8.6} & 1.48 & 0.15 & 0.90 & 0.12 & & &  \\
\textbf{5.75} & 5.70 & 0.56 & \textbf{9.9} & 1.32 & 0.13 & 0.760 & 0.099 & & &  \\
\textbf{6.30} & 5.05 & 0.46 & \textbf{11.1} & 1.08 & 0.11 & 0.594 & 0.082 & & &  \\
\textbf{6.90} & 3.77 & 0.38 & \textbf{12.4} & 0.645 & 0.081 & 0.717 & 0.086 & & &  \\
\textbf{7.50} & 3.36 & 0.34 & \textbf{13.8} & 0.653 & 0.071 & 0.530 & 0.064 & & &  \\
\textbf{8.20} & 2.47 & 0.24 & \textbf{15.2} & 0.582 & 0.067 & 0.431 & 0.058 & & &  \\
\textbf{8.90} & 3.10 & 0.30 & \textbf{17.0} & 0.438 & 0.051 & 0.434 & 0.051 & & &  \\
\textbf{9.60} & 2.71 & 0.24 & \textbf{20.5} & 0.362 & 0.046 & 0.502 & 0.054 & & &  \\
\textbf{10.5} & 1.83 & 0.17 & & & & & & & & \\
\textbf{11.5} & 2.10 & 0.17 & & & & & & & & \\
\textbf{12.5} & 1.74 & 0.15 & & & & & & & & \\
\textbf{13.5} & 1.67 & 0.14 & & & & & & & & \\
\textbf{15.0} & 1.49 & 0.10 & & & & & & & & \\
\textbf{17.2} & 1.274 & 0.087 & & & & & & & & \\
\textbf{20.2} & 1.033 & 0.093 & & & & & & & & \\ \hline
\end{tabular}
\caption{GC radial surface density data for the total GC system and each subpopulation. We show surface density with different radial binning (columns typeset in bold) for the total GC system, the blue and red subpopulations and the green subpopulation.} %Galactocentric radii ($r_{gal}$) are quoted in arcmin, surface density ($\rho$) and error in surface density ($\epsilon_{\rho}$) are quoted in objects per arcmin$^-2$.}
\label{tab:sd}
\end{table*}

\end{appendix}

\end{document}